%% file: main.tex
\documentclass[a4paper,12pt]{article}
\pdfoutput=1
\usepackage[top=2.5cm, bottom=2.5cm, outer=1.5 cm, inner=1.5cm]{geometry}
\usepackage{graphicx}
\usepackage{epsfig}
\usepackage{amsfonts,amssymb,amsmath,mathrsfs}
\usepackage{cite}
\usepackage{dsfont}
\usepackage{float}
\usepackage[font=footnotesize,labelfont=bf,justification=centerlast,width=.94\textwidth]{caption}
\usepackage{authblk}
\usepackage{comment}
\usepackage{hyperref}
\usepackage[utf8]{inputenc}
\usepackage[toc,page]{appendix}
\usepackage{color,xcolor}
\usepackage{cancel}
\usepackage{tikz}
\usepackage{comment}
\usetikzlibrary{automata,positioning,arrows}
\usetikzlibrary{shapes,shadows,arrows}

\numberwithin{equation}{section}
\begin{document}

\include{macros}

\title{\textbf{Fractional Cosmic String Loops In Expanding Universe }}

\author[a]{Pankaj Chaturvedi \thanks{\noindent E-mail:~  cpankaj1@gmail.com}}
\author[a]{Bikram Nath \thanks{\noindent E-mail:~ bikram24\_rs@phy.nits.ac.in}}

\affil[a]{
\textit{Department of Physics, NIT Silchar, Assam  788010, India}}

\maketitle

\abstract{
We study the dynamics of circular cosmic string loops in a spatially flat Friedmann Lemaître Robertson Walker universe within a fractional Polyakov framework that incorporates nonlocal memory effects. Allowing both the loop radius and polar angle to evolve, we obtain a coupled non-autonomous system governed by string tension, cosmological expansion, and an emergent centrifugal contribution. We show that angular dynamics plays a crucial role in determining the loop evolution. In contrast to standard scenarios where loops collapse, we identify a class of solutions exhibiting sustained expansion driven by dynamically generated angular motion. The system also displays nonlinear behavior with signatures of chaos, with the onset of chaotic dynamics closely correlated with expanding solutions. Our results demonstrate that fractional memory effects and angular degrees of freedom qualitatively modify cosmic string loop dynamics, providing new mechanisms for stability in cosmological backgrounds.
}

\newpage
\tableofcontents
\addtocontents{toc}{\protect\setcounter{tocdepth}{2}}
\setcounter{page}{1}


\section{Introduction}

Cosmic strings are one-dimensional topological defects that may have formed during symmetry-breaking phase transitions in the early universe \cite{Kibble1976}. Their existence is closely tied to the topology of the vacuum manifold of the underlying quantum field theory. In particular, when a continuous symmetry is spontaneously broken and the vacuum manifold possesses a nontrivial first homotopy group, stable line-like configurations can emerge. These objects are of considerable interest because they provide a rare connection between high-energy particle physics and cosmology, offering a possible observational window into physical processes that took place at extremely high energy scales shortly after the Big Bang.

The physical properties of cosmic strings have been studied in detail over several decades \cite{Vilenkin1981,VilenkinShellard1994,HindmarshKibble1995}. In the thin-string approximation, they can be described as relativistic one-dimensional objects with a large energy per unit length, or tension, denoted by $\mu$. Their dynamics is governed by the Nambu--Goto action, or equivalently by the Polyakov action, both of which describe the evolution of a string worldsheet embedded in spacetime. The gravitational influence of cosmic strings is determined by the dimensionless parameter $G\mu$, where $G$ is Newton’s constant. Observational constraints indicate that this parameter is small, yet potentially within the range relevant for cosmological signatures. In recent years, data from pulsar timing arrays such as NANOGrav, EPTA, and PPTA have provided increasingly precise constraints on cosmic string models, while also keeping open the intriguing possibility that a stochastic gravitational wave background may contain contributions from string networks \cite{Auclair2020,Blasi2021,Sousa:2024ytl}.

Cosmic strings arise in a wide range of theoretical settings. In classical field theory, they appear as vortex solutions similar to the Nielsen–Olesen string in the Abelian Higgs model \cite{NielsenOlesen1973}. More generally, they occur whenever the topology of the vacuum manifold allows for stable one-dimensional defects \cite{Vachaspati1991}. Depending on the underlying symmetry-breaking pattern, cosmic strings may be classified as local, global, superconducting, or semilocal \cite{Witten1985,AchucarroVachaspati2000}. In addition, developments in string theory have introduced the concept of cosmic superstrings, which can form in brane inflation scenarios and exhibit a rich phenomenology that differs in important ways from field-theoretic strings \cite{CopelandMyersPolchinski2004,SarangiTye2002}.

After their formation, cosmic strings evolve into networks consisting of long strings that span cosmological distances and a population of closed loops. A combination of analytical arguments and numerical simulations has shown that these networks tend toward a scaling regime in which their statistical properties evolve in proportion to the size of the cosmological horizon \cite{AlbrechtTurok1989,BennettBouchet1988,AllenShellard1990}. In this regime, loop production plays a central role. Loops are continuously formed through the intersection and reconnection of strings, and they provide an efficient mechanism through which the network loses energy \cite{PolchinskiRocha2007}.

Cosmic string loops are therefore of particular importance for both theoretical studies and observational searches. Once formed, loops undergo relativistic oscillations and emit gravitational radiation, gradually losing energy over time \cite{VachaspatiVilenkin1985,DamourVilenkin2000,DamourVilenkin2005}. The gravitational waves produced by these oscillations span a wide range of frequencies and are actively searched for by pulsar timing arrays, ground-based detectors, and future space-based missions \cite{Siemens2006,Olmez2010,Auclair2020}. In flat spacetime, the motion of loops is well understood and is characterized by periodic behavior, including the formation of cusps and kinks. In an expanding universe, however, the situation becomes more subtle. The evolution of a loop is influenced by the competition between its intrinsic tension, which tends to contract it, and the expansion of the universe, which tends to stretch it \cite{VilenkinShellard1994}. Under certain conditions, such as in accelerating backgrounds, sufficiently large loops may even expand instead of collapsing \cite{Larsen1994}. Additional modifications, including time-dependent tension or interactions with external fields, can further enrich the dynamics \cite{Yamaguchi2006,LiuCheng2008}. Nevertheless, within the standard framework, loop evolution is largely governed by local worldsheet physics and typically results in eventual collapse.

An important assumption underlying the conventional description of cosmic strings is that the dynamics is local on the worldsheet. This means that the evolution at a given point depends only on the fields and their derivatives evaluated at that point. While this approximation has been extremely successful, there are strong reasons to consider more general frameworks in which nonlocal effects and memory-dependent interactions are present. Such effects naturally arise in many areas of physics, including complex systems, effective field theories, and certain approaches to quantum gravity. Fractional calculus provides a powerful and mathematically consistent way to incorporate such features by allowing derivatives and integrals of non-integer order \cite{Podlubny1999,Klafter2011,Calcagni2010}. This formalism has been successfully applied to describe systems with memory, dissipation, and anomalous transport behavior.

Motivated by these developments, fractional extensions of classical and quantum theories have been explored in recent years. In the context of string theory, fractional generalizations of the Polyakov action have been proposed in which the standard worldsheet measure is modified to include a fractional kernel \cite{Calcagni2010,Diaz2018}. As a result, the equations of motion acquire additional terms that reflect the nonlocal nature of the dynamics. These terms often resemble time-dependent dissipative contributions and encode the influence of the system’s past evolution. The fractional parameter controls the strength of these effects and allows one to interpolate continuously between standard local dynamics and nonlocal behavior.

Despite the conceptual appeal of fractional string models, their implications for cosmic string dynamics have not yet been fully explored. In particular, the evolution of cosmic string loops in a cosmological background, when both memory effects and additional degrees of freedom are taken into account, remains largely uncharted. This is an important gap, since even small modifications to loop dynamics can lead to significant changes in their lifetime, stability, and observational signatures.

In this work, we investigate the dynamics of fractional cosmic string loops in a spatially flat Friedmann–Lemaître–Robertson–Walker universe. Starting from a fractional Polyakov action, we derive the effective equations governing circular loop configurations. A key aspect of our approach is the inclusion of angular dynamics on the two-sphere, which introduces an additional degree of freedom beyond the usual radial motion and leads to a richer dynamical system.

Our analysis reveals several interesting and previously unexplored features. In particular, we find a class of solutions in which cosmic string loops can undergo sustained expansion driven by dynamically generated angular motion. This behavior contrasts with the conventional expectation that loops eventually collapse in an expanding universe. In addition, the system exhibits nonlinear dynamics with sensitivity to initial conditions, suggesting the presence of chaotic behavior in certain regimes. These results indicate that fractional memory effects, when combined with angular degrees of freedom, can qualitatively alter the evolution of cosmic string loops and open new possibilities for their long-term survival in cosmological settings.

The paper is organized as follows. In Section 2 we review the fractional action-like variational formalism and its application to bosonic strings. In Section 3 we study the dynamics of fractional cosmic string loops in a flat FLRW background. We first analyze the case of fixed polar angle and derive the corresponding effective dynamics. We then consider the more general case in which the polar angle is dynamical, leading to a coupled system with both radial and angular degrees of freedom. We investigate the resulting dynamics numerically and analyze the emergence of expanding solutions and chaotic behavior. Finally, we summarize our results and discuss their implications.
\section{Fractional Action-Like Variational Problems}

Fractional action-like variational problems provide an alternative formulation of dynamics in which nonlocality is incorporated directly at the level of the action functional. Unlike conventional approaches to fractional mechanics, where fractional derivatives are introduced into the equations of motion, the fractional action-like variational approach modifies the action itself through a fractional integral. This framework was developed by El-Nabulsi and collaborators as a systematic method for describing nonconservative systems within a variational setting \cite{ElNabulsi2005,ElNabulsi2005b}. A detailed mathematical formulation and further developments can be found in \cite{ElNabulsiTorres2008}.

A characteristic feature of this formalism is the presence of two distinct time parameters. One distinguishes between an intrinsic time variable $\tau$, which parametrizes the trajectory, and an observer time $t$, which determines the upper limit of integration. This separation reflects the fundamentally nonlocal nature of the dynamics, since the evolution depends on a weighted history of the system rather than on instantaneous values alone \cite{Udriste2005,Udriste2007}. In its simplest one-dimensional form, the fractional action-like functional is defined as
\begin{equation}
S_{t_{0}+}^\alpha[q](t)
=
\frac{1}{\Gamma(\alpha)}
\int_{t_{0}}^t
L\big(\dot{q}(\tau),q(\tau),\tau\big)
(t-\tau)^{\alpha-1} d\tau ,
\label{falva_action}
\end{equation}
where $0<\alpha<1$ and $\Gamma(\alpha)$ denotes the Euler gamma function. The kernel $(t-\tau)^{\alpha-1}$ weights past contributions to the action and introduces a memory effect. In the limit $\alpha \to 1$, the kernel reduces to unity and \eqref{falva_action} recovers the standard action of classical mechanics.

An important aspect of \eqref{falva_action} is that the fractional structure is entirely encoded in the integration measure. As a consequence, the resulting equations of motion retain a form closely related to the classical Euler–Lagrange equations, but acquire additional contributions that reflect the nonlocality of the dynamics. Performing a variation of \eqref{falva_action} with fixed endpoints yields the fractional Euler–Lagrange equations
\begin{equation}
\frac{\partial L}{\partial q^i}
-
\frac{d}{d\tau}
\left(
\frac{\partial L}{\partial \dot{q}^i}
\right)
=
\frac{1-\alpha}{t-\tau}
\frac{\partial L}{\partial \dot{q}^i},
\label{falva_eom}
\end{equation}
where all quantities are evaluated along the trajectory $q(\tau)$. The additional term on the right-hand side originates from the fractional weight in \eqref{falva_action} and has no analogue in the standard variational principle. It is useful to rewrite \eqref{falva_eom} in a form that highlights its physical interpretation. Introducing the Euler–Lagrange operator
\begin{equation}
E(L)=\frac{\partial L}{\partial q}-\frac{d}{d\tau}\frac{\partial L}{\partial \dot{q}},
\end{equation}
one can express \eqref{falva_eom} as
\begin{equation}
E(L)=\frac{\partial Q}{\partial \dot{q}}, \qquad
Q=\frac{1-\alpha}{t-\tau}L ,
\label{falva_rayleigh_form}
\end{equation}
where the function $Q$ plays the role of an effective Rayleigh dissipation function. This representation shows that the fractional action-like formalism naturally incorporates dissipative effects while preserving a variational origin. The dissipation is not introduced phenomenologically but emerges directly from the nonlocal structure of the action.

The formalism can be extended by allowing the Lagrangian to depend on fractional derivatives of the dynamical variables. In this case one replaces the ordinary derivative $\dot{q}$ in \eqref{falva_action} by a fractional derivative operator. A particularly useful choice is the operator introduced by Cresson, which combines left and right Riemann–Liouville derivatives \cite{Cresson2007,ElNabulsiTorres2008}. The generalized functional then takes the form
\begin{equation}
S^{\alpha,\beta}_\gamma[q](t)
=
\frac{1}{\Gamma(\alpha)}
\int_{t_0}^t
L\big(D^{\alpha,\beta}_\gamma q(\tau),q(\tau),\tau\big)
(t-\tau)^{\alpha-1} d\tau ,
\label{falva_fractional_derivative_action}
\end{equation}
where the operator $D^{\alpha,\beta}_\gamma$ in the above fractional action is defined as
\begin{equation}
D^{\alpha,\beta}_{\gamma}
=
\frac{1}{2}\left(D^{\alpha}_{t_{0}+}-D^{\beta}_{t-}\right)
+
\frac{i\gamma}{2}\left(D^{\alpha}_{t_{0}+}+D^{\beta}_{t-}\right),
\label{cresson_derivative}
\end{equation}
with $\gamma \in \mathbb{C}$ and the derivatives $D^{\alpha}_{t_{0}+}$ and $D^{\beta}_{t-}$ are defined as the left and right Riemann–Liouville fractional derivatives of order $0<\alpha,\beta<1$, given by
\begin{align}
D^{\alpha}_{t_{0}+} f(\theta)
&=
\frac{1}{\Gamma(1-\alpha)}
\frac{d}{d\theta}
\int_a^\theta
f(\tau)(\theta-\tau)^{-\alpha} d\tau, \\
D^{\beta}_{t-} f(\theta)
&=
\frac{1}{\Gamma(1-\beta)}
\left(-\frac{d}{d\theta}\right)
\int_\theta^t
f(\tau)(\tau-\theta)^{-\beta} d\tau .
\end{align}

The construction described above interpolates between purely past-dependent and future-dependent derivatives. For $\gamma=-i$ one recovers the left derivative $D^{\alpha}_{t_{0}+}$, while for $\gamma=i$ one obtains $-D^{\beta}_{t-}$. In the limit $\alpha \to 1$ and $\beta \to 1$, the operator reduces to the ordinary derivative $d/d\tau$. The presence of both left and right derivatives reflects the intrinsic nonlocality of the theory: the left derivative encodes dependence on past history, while the right derivative introduces sensitivity to future boundary conditions. The parameter $\gamma$ controls the relative contribution of these effects. The corresponding Euler–Lagrange equations generalize \eqref{falva_eom} and involve fractional derivatives acting on the generalized momenta \cite{ElNabulsiTorres2008}. However, an important structural feature remains unchanged: the term proportional to $(1-\alpha)/(t-\tau)$ persists in the equations of motion. This shows that the effective dissipative contribution is not tied to the specific choice of derivative operator but is instead a universal consequence of the fractional measure in the action.

From a physical point of view, the fractional action-like formalism provides a unified framework for incorporating memory and dissipation into variational principles. The equations of motion derived from \eqref{falva_action} and \eqref{falva_fractional_derivative_action} describe systems whose evolution deviates from conservative dynamics in a controlled manner governed by the fractional parameters. At the same time, the variational origin ensures that the structure of the theory remains close to classical mechanics. Despite these appealing features, several foundational issues remain open. In particular, the existence of extremals for functionals of the form \eqref{falva_action} is not guaranteed in general \cite{ElNabulsiTorres2008}. This indicates that the mathematical foundations of fractional variational principles are not yet fully understood and require further investigation.

In applications to string dynamics and cosmology, the formalism is especially suggestive. The presence of intrinsic nonlocality and time-dependent dissipative terms indicates that fractional action principles may provide an effective description of systems evolving in complex backgrounds where standard local dynamics is insufficient. In this sense, the fractional action-like approach offers a natural extension of the classical action principle, capable of capturing a broader class of physical phenomena.

\section{Fractional Bosonic Strings}

The fractional action-like variational framework discussed in the previous section provides a natural starting point for constructing fractional extensions of string dynamics. In particular, the Polyakov formulation of bosonic strings can be interpreted as a two-dimensional field theory whose action is defined on the worldsheet. It is therefore natural to view fractional bosonic strings as a special case of fractional action-like variational problems in which the configuration variables are the embedding fields $X^\mu(\tau,\sigma)$ and the fractional structure is introduced through a nontrivial integration measure along the worldsheet coordinates.

In standard bosonic string theory, the worldsheet dynamics is governed by the Polyakov action
\begin{equation}
S_P=
-\frac{1}{4\pi\alpha'}
\int d\tau\, d\sigma\,
\sqrt{-h}\, h^{ab}
\,\partial_a X \cdot \partial_b X ,
\label{Polyakov}
\end{equation}
which is invariant under worldsheet diffeomorphisms and Weyl transformations \cite{Nambu1970,Goto1971,VilenkinShellard1994}. Within the fractional action-like framework, the natural generalization is obtained by replacing the standard measure $d\tau$ with the fractional kernel appearing in \eqref{falva_action}. Restricting to the case in which the fractional structure is introduced only along the temporal direction of the worldsheet, the action takes the form
\begin{equation}
S_{t_{0}+}^\alpha =
-\frac{1}{4\pi\alpha'\Gamma(\alpha)}
\int_0^\ell d\sigma
\int_{t_0}^{t}
(t-\tau)^{\alpha-1} d\tau
\sqrt{-h}\,h^{ab}
\,\partial_a X \cdot \partial_b X .
\label{timefractionalaction}
\end{equation}
This expression can be viewed as a direct realization of the functional \eqref{falva_action} with the Lagrangian density identified as the Polyakov Lagrangian. Considering the limits in the order $\alpha \to 1$, $t\to\infty$, and $t_0\to -\infty$ the kernel becomes trivial and \eqref{timefractionalaction} reduces to the standard Polyakov action \eqref{Polyakov}, showing that the fractional theory provides a continuous deformation of the conventional string dynamics \cite{Diaz2018}.

Varying the action \eqref{timefractionalaction} with respect to the embedding coordinates $X^\mu$ yields the equations of motion
\begin{equation}
\partial_a\left[
v_\alpha(\tau)\sqrt{-h}\,h^{ab}\partial_b X^\mu
\right]=0,
\label{fractionalstringeom}
\end{equation}
where $v_\alpha(\tau)=(t-\tau)^{\alpha-1}/\Gamma(\alpha)$. This equation is the direct analogue of the fractional Euler–Lagrange equation \eqref{falva_eom} applied to a two-dimensional field theory. Expanding the derivative, one finds
\begin{equation}
\Box X^\mu + \frac{\partial_\tau v_\alpha}{v_\alpha}\,\partial_\tau X^\mu =0,
\end{equation}
where $\Box$ denotes the standard worldsheet d'Alembertian. The second term introduces an explicit time-dependent contribution proportional to $(1-\alpha)/(t-\tau)$, which plays the role of a damping term and reflects the memory effects inherent in the fractional action.

Variation with respect to the worldsheet metric $h_{ab}$ leads to the stress-energy tensor
\begin{equation}
T_{ab}
=
\frac{1}{\alpha'}
\left(
\partial_a X\cdot\partial_b X
-
\frac12 h_{ab}h^{cd}\partial_c X\cdot\partial_d X
\right),
\label{stressenergy}
\end{equation}
and the classical constraints $T_{ab}=0$ still arise as equations of motion. However, unlike the standard case, the presence of the fractional measure implies that the stress tensor is not conserved. Taking the covariant divergence of \eqref{stressenergy} and using \eqref{fractionalstringeom}, one obtains
\begin{equation}
\nabla^a T_{ab}
=
-\frac{\partial_b v_\alpha}{v_\alpha}\,\mathcal{L},
\end{equation}
where $\mathcal{L}$ is the Polyakov Lagrangian density. This shows that the worldsheet energy-momentum is not conserved locally but instead exchanges with an effective external sector induced by the fractional kernel. This behavior is the direct field theoretic analogue of the dissipative term appearing in \eqref{falva_eom}.

\subsection{Symmetry breaking and constraint structure}

The introduction of the fractional kernel in \eqref{timefractionalaction} has profound consequences for the symmetry structure of the worldsheet theory. In the standard Polyakov formulation, invariance under arbitrary reparametrizations $(\tau,\sigma)\to(\tau',\sigma')$ allows one to impose the conformal gauge $h_{ab}=\eta_{ab}$ and leads to the Virasoro constraints generating a conformal symmetry algebra. In the fractional theory, however, the explicit dependence of $v_\alpha(\tau)$ on $\tau$ breaks reparametrization invariance in the temporal direction. Spatial reparametrizations $\sigma\to\sigma'(\sigma)$ and Weyl rescalings remain intact, but the full two-dimensional diffeomorphism invariance is lost. As a consequence, the worldsheet metric cannot be completely gauge-fixed to the flat form. A convenient partial gauge fixing is
\begin{equation}
h_{ab}=
\begin{pmatrix}
-1 & 0 \\
0 & \omega^2(\tau,\sigma)
\end{pmatrix},\label{wmetric}
\end{equation}
where $\omega$ encodes the residual degree of freedom associated with the reduced symmetry. In this gauge, the constraints $T_{ab}=0$ take the form
\begin{align}
T_{\tau\tau} &= \frac{1}{2\alpha'}\left(\dot{X}^2+\omega^2 X'^2\right)=0, \\
T_{\tau\sigma} &= \frac{1}{\alpha'}\dot{X}\cdot X'=0.
\end{align}

These relations generalize the usual Virasoro constraints, but now involve the function $\omega(\tau,\sigma)$, which cannot be eliminated by gauge fixing. Thus the constraints acquire explicit dependence on the fractional dynamics. The Hamiltonian formulation further clarifies the structure of the theory. The canonical momentum conjugate to $X^\mu$ is
\begin{equation}
P_\mu =
\frac{v_\alpha(\tau)}{2\pi\alpha'}
\sqrt{-h}\,h^{\tau a}\partial_a X_\mu,
\end{equation}
which differs from the standard expression by the factor $v_\alpha(\tau)$. The Hamiltonian density becomes
\begin{equation}
\mathcal{H}
=
\frac{1}{4\pi\alpha' v_\alpha(\tau)}
\left[
P^2 + v_\alpha^2(\tau)\,X'^2
\right].
\end{equation}
The primary constraints can be written as
\begin{equation}
\mathcal{H}_\perp = \frac{1}{4\pi\alpha'}\left(\frac{P^2}{v_\alpha}+v_\alpha X'^2\right)\approx0,
\qquad
\mathcal{H}_\parallel = \frac{1}{2\pi\alpha'} P\cdot X' \approx 0.
\end{equation}

The Poisson brackets of these constraints reveal that the algebra is deformed. In particular, one finds schematically
\begin{align}
\{\mathcal{H}_\perp(\sigma),\mathcal{H}_\perp(\sigma')\}
&\sim
v_\alpha(\tau)\,\mathcal{H}_\parallel\,\partial_\sigma\delta(\sigma-\sigma'), \\
\{\mathcal{H}_\parallel(\sigma),\mathcal{H}_\perp(\sigma')\}
&\sim
\mathcal{H}_\perp\,\partial_\sigma\delta(\sigma-\sigma').
\end{align}
The presence of the explicit factor $v_\alpha(\tau)$ indicates that the structure constants of the algebra are replaced by structure functions depending on $\tau$. Consequently, the usual Virasoro algebra does not survive in its standard form. Instead, one obtains a deformed algebra reflecting the breaking of time reparametrization invariance.

This deformation has important implications. Since the Virasoro algebra encodes conformal symmetry, its modification implies that the fractional string does not define a conventional conformal field theory. The absence of full conformal invariance affects the gauge fixing, the spectrum, and potentially the consistency conditions of the theory. At the classical level, the equations of motion already exhibit dissipative behavior, and the worldsheet dynamics incorporates memory effects inherited from the fractional action. These features distinguish fractional bosonic strings from their conventional counterparts and make them a natural framework for exploring string dynamics in nonlocal or cosmological settings.
\section{Fractional cosmic string loops in an expanding universe}
The formalism developed above provides a consistent framework for studying string dynamics in the presence of nonlocal and dissipative effects. We now apply this fractional string description to the evolution of cosmic string loops in an expanding cosmological background. On sufficiently large scales the universe is well described by the FLRW spacetime. In comoving coordinates the line element is
\begin{equation}
ds^2=-d\tau^2+a^2(\tau)\left[d\chi^2+\Sigma_K^2(\chi)d\theta^2+\Sigma_K^2(\chi)\sin^2\theta\,d\phi^2\right],
\label{FLRWmetric}
\end{equation}
where $a(\tau)$ is the cosmological scale factor and  
$\Sigma_K(\chi)=\{\sin\chi,\chi,\sinh\chi\}$ corresponds to spatial curvature
$K=\{1,0,-1\}$ respectively \cite{VilenkinShellard1994,HindmarshKibble1995}.  
In the present work we restrict our attention to the spatially flat case $K=0$, for which $\Sigma_K(\chi)=\chi$. Equation \eqref{FLRWmetric} then reduces to the standard flat FLRW geometry that is widely used in cosmological applications.

The time dependence of the scale factor depends on the dominant component of the cosmic energy density. In particular one has

\begin{equation}
a(t)\propto
\begin{cases}
\tau^{1/2}, & \text{radiation dominated era}, \\
\tau^{2/3}, & \text{matter dominated era}, \\
e^{H\tau}, & \text{de Sitter phase}.
\end{cases}
\label{scalefactor}
\end{equation}

These expressions reflect the successive stages of cosmological evolution. After an early inflationary period the universe reheated and entered a hot radiation dominated phase in which relativistic particles dominated the energy density. This radiation era extends approximately from $\tau\sim10^{-32}\,\mathrm{s}$ until $\tau\sim10^{12}\,\mathrm{s}$, after which the universe gradually becomes matter dominated. At much later times the expansion becomes accelerated due to dark energy.

Cosmic strings produced during early symmetry breaking phase transitions would begin their dynamical evolution during the radiation dominated epoch. In this regime the evolution of a string loop is governed by the competition between two effects. The intrinsic string tension tends to contract the loop, while the expansion of the universe tends to stretch it. The resulting dynamics therefore depends both on the properties of the string and on the cosmological background in which it evolves \cite{VilenkinShellard1994}. For this reason the radiation dominated era provides a natural setting for studying the early behaviour of cosmic string loops.

In order to explore possible modifications of this dynamics we consider a fractional generalization of the string worldsheet action. Following the fractional string formulation proposed in \cite{Diaz2018},  we consider the fractional Polyakov action introduced in \eqref{timefractionalaction}, now embedded in the curved FLRW background. Throughout this work we adopt the convenient normalization $2\pi\alpha'=1$. The resulting time-fractional Polyakov action can be written as
\begin{equation}
S^\alpha =
-\frac{1}{2\,\Gamma(\alpha)}
\int_0^\ell d\sigma^1
\int_{-\infty}^{t}
(t-\sigma^0)^{\alpha-1} d\sigma^0
\sqrt{-h}\,h^{ab}
\,g_{\mu\nu}(X)\,\partial_a X^\mu \partial_b X^\nu ,
\label{FractionalAction}
\end{equation}
where $\sigma^0$ and $\sigma^1$ denote the worldsheet time and spatial coordinates respectively, $h_{ab}$ is the intrinsic worldsheet metric, and $g_{\mu\nu}(X)$ is the background spacetime metric given by Eq.~\eqref{FLRWmetric}. The parameter $\alpha$ controls the strength of the fractional deformation and interpolates between fractional and ordinary string dynamics. In the successive limits $\alpha\rightarrow 1$ and $t\to\infty$ the fractional kernel in Eq.~\eqref{FractionalAction} reduces to the standard integration measure and the usual Polyakov action is recovered \cite{Diaz2018}.

The parameter $t$ specifies the upper limit of the fractional time integration and determines the temporal scale over which the worldsheet history contributes to the nonlocal dynamics. Physically, this parameter controls the effective memory length of the system. Since we are interested in the behaviour of cosmic string loops during the radiation dominated epoch, we choose $t$ to correspond approximately to the cosmic time at which the radiation era ends and the universe transitions to matter domination. In our numerical analysis we therefore take $t \simeq 1.6\times10^{12}\,\mathrm{sec}$, which is of the order of the radiation–matter equality time.

The dynamics of a cosmic string loop is obtained by specifying the embedding of the worldsheet into the FLRW spacetime described by Eq.~\eqref{FLRWmetric}. The equations of motion follow from varying the action \eqref{FractionalAction} with respect to the embedding coordinates $X^\mu$ and the worldsheet metric $h_{ab}$. As discussed in the previous section, the fractional measure introduces an explicit time dependence into the equations of motion, leading to modified evolution equations that contain damping-like terms proportional to $(1-\alpha)/(t-\sigma^0)$. These terms encode the memory effects associated with the fractional dynamics and can significantly alter the behaviour of string loops compared to the standard case. In the following subsections, we analyze two physically relevant configurations for the string embedding and derive the corresponding equations governing the evolution of fractional cosmic string loops in the radiation-dominated universe.

\subsection{Circular loop with fixed polar angle}

We first consider the dynamics of a circular cosmic string loop confined to the equatorial plane of the FLRW spacetime. In this configuration, the embedding of the string worldsheet into the background geometry given by Eq.~\eqref{FLRWmetric} is chosen as

\begin{equation}
\sigma^0=\tau,\qquad
\chi=\chi(\tau),\qquad
\theta=\theta_0,\qquad
\sigma^1=\phi .
\label{embedding1}
\end{equation}
where $\tau$ represents the worldsheet time coordinate, while $\phi$ parameterizes the spatial direction along the string. This embedding describes a circular loop whose comoving radius $\chi(\tau)$ evolves with cosmic time. The parameter $\theta_0$ can take values in the interval $[0,\pi/2]$. Owing to the spherical symmetry of the FLRW geometry, this restriction does not reduce the generality of the physical description.

For the spatially flat universe ($K=0$) considered here, one has $\Sigma_K(\chi)=\chi$. Moreover, during the radiation dominated epoch, the cosmological expansion is governed by the scale factor
\begin{equation}
a(\tau)=\sqrt{2a_0^2 H_0 \tau},
\label{radiationscale}
\end{equation}
which gives the Hubble parameter as $H=\dot{a}/a=1/(2\tau)$, a relation that will play an important role in simplifying the dynamics below. It is also convenient to introduce the physical radius of the loop,
\begin{equation}
R(\tau)=a(\tau)\chi(\tau),
\label{physicalradius}
\end{equation}
which measures the proper size of the loop in the expanding universe. Expressing the dynamics in terms of $R(\tau)$ allows one to separate the intrinsic contraction due to string tension from the background cosmological expansion.

We now derive the effective dynamics starting directly from the fractional Polyakov action \eqref{FractionalAction}. As discussed in the previous section, the breaking of time reparametrization invariance prevents a complete conformal gauge fixing, and the worldsheet metric can be written in the partially gauge-fixed form as given in eq.\eqref{wmetric}. However, the circular loop ansatz \eqref{embedding1} significantly simplifies the structure of the induced geometry. Since $\partial_\sigma \chi=0$, the mixed component of the induced metric vanishes identically,
\begin{equation}
\gamma_{\tau\sigma}
=
g_{\mu\nu}\,\partial_\tau X^\mu \partial_\sigma X^\nu
=0,
\label{gamma_tausigma}
\end{equation}
which is consistent with the constraint $T_{\tau\sigma}=0$. Furthermore, the $\sigma\sigma$ component becomes
\begin{equation}
\gamma_{\sigma\sigma}
=
g_{\mu\nu}\,\partial_\sigma X^\mu \partial_\sigma X^\nu
=
a^2(\tau)\chi^2(\tau)\sin^2\theta_0,
\label{gamma_sigmasigma}
\end{equation}
and depends only on $\tau$. As a consequence, the residual degree of freedom $\omega(\tau,\sigma)$ can be consistently taken to depend only on $\tau$. Making use of the remaining Weyl invariance of the worldsheet theory, one can then fix this function to a constant without loss of generality. In particular, choosing $\omega(\tau,\sigma)=1,$
brings the worldsheet metric to the simple form
\begin{equation}
h_{ab}=\eta_{ab}=\mathrm{diag}(-1,1).
\label{metric_flat}
\end{equation}

Thus, although full conformal symmetry is not available in the fractional theory, the symmetry of the circular loop configuration effectively allows a reduction to a flat worldsheet metric. Now, substituting the embedding \eqref{embedding1} together with the gauge-fixed metric \eqref{metric_flat} into the fractional Polyakov action \eqref{FractionalAction} and using the relations \eqref{radiationscale} and \eqref{physicalradius}, one obtains the effective Lagrangian density governing the loop dynamics as
\begin{equation}
\mathcal{L}
=
-\frac{(t-\tau)^{\alpha-1}}{2\Gamma(\alpha)}
\left[
1
- \left(\dot{R}-\frac{R}{2\tau}\right)^2
- R^2\sin^2\theta_0
\right].
\label{lagrangian_R}
\end{equation}

The structure of the Lagrangian given above makes the physical content transparent. The term $(\dot{R}-R/2\tau)$ reflects the interplay between intrinsic string dynamics and cosmological expansion, while the fractional prefactor $(t-\tau)^{\alpha-1}$ introduces a time-dependent weighting that encodes memory effects. In the limit $\alpha\to1$, the standard Polyakov dynamics in an expanding background is recovered, whereas for $\alpha<1$ the evolution acquires dissipative features characteristic of fractional systems.

We will now determine the Hamiltonian formulation of the system in order to study the loop dynamics in a more transparent way. The canonical momentum conjugate to the loop radius $R(\tau)$ is defined as $P_R(\tau)=\partial\mathcal{L}/\partial \dot{R}(\tau)$. Using the Lagrangian \eqref{lagrangian_R}, it can be given explicitly as
\begin{equation}
P_R(\tau)
=
\frac{(t-\tau)^{\alpha-1}}{\Gamma(\alpha)}
\left(\dot{R}-\frac{R}{2\tau}\right).
\label{momentum_R}
\end{equation}

This expression shows that the canonical momentum is naturally associated with the shifted velocity $(\dot{R}-R/2\tau)$ rather than $\dot{R}$ itself. The shift originates from the cosmological expansion and reflects the fact that $R(\tau)$ already contains the scale factor. The prefactor $(t-\tau)^{\alpha-1}$ introduces a time-dependent weighting which encodes the nonlocal memory effects characteristic of the fractional dynamics.

The Hamiltonian follows from the Legendre transformation $\mathcal{H}=P_R\dot{R}-\mathcal{L}$. Eliminating $\dot{R}$ in favour of $P_R$ using \eqref{momentum_R}, one obtains
\begin{equation}
\mathcal{H}
=
\frac{\Gamma(\alpha)}{2\,(t-\tau)^{\alpha-1}}\,P_R^2
+
\frac{P_R R}{2\tau}
+
\frac{(t-\tau)^{\alpha-1}}{2\Gamma(\alpha)}
\left[
1 - R^2\sin^2\theta_0
\right].
\label{Hamiltonian_R}
\end{equation}

The structure of the Hamiltonian makes the physical content particularly transparent. The first term represents the kinetic contribution expressed in terms of the canonical momentum, the second term arises purely from the cosmological expansion and couples $R$ and $P_R$, while the last term plays the role of an effective potential determined by the string tension and the fractional kernel. The explicit dependence on $\tau$ indicates that the system is non-autonomous, reflecting both the expanding background and the time-dependent fractional measure. Hamilton’s equations then take the form
\begin{equation}
\dot{R}
=
\frac{\partial \mathcal{H}}{\partial P_R}
=
\frac{\Gamma(\alpha)}{(t-\tau)^{\alpha-1}}\,P_R
+
\frac{R}{2\tau},
\label{Rdot_eq}
\end{equation}
\begin{equation}
\dot{P}_R
=
-\,\frac{\partial \mathcal{H}}{\partial R}
=
-\frac{P_R}{2\tau}
+
\frac{(t-\tau)^{\alpha-1}}{\Gamma(\alpha)}\,R\sin^2\theta_0 .
\label{Pdot_eq}
\end{equation}

These equations form a closed dynamical system governing the evolution of the loop radius. The terms proportional to $1/\tau$ encode the effect of cosmological expansion and act as damping (or dilution) terms, while the terms proportional to $R\sin^2\theta_0$ represent the restoring force due to the string tension. The fractional parameter $\alpha$ modifies both contributions through a time-dependent coupling, thereby incorporating memory effects into the dynamics. To ensure a consistent evolution, the initial data must satisfy the worldsheet constraint $T_{\tau\tau}=0$. Using
\begin{equation}
T_{\tau\tau}
=
\frac{1}{2\alpha'}
\left[
\dot{X}^2 + X'^2
\right]=0,
\end{equation}
and evaluating it for the circular loop ansatz \eqref{embedding1}, we obtain
\begin{equation}
1
-
\left(\dot{R}-\frac{R}{2\tau}\right)^2
-
R^2\sin^2\theta_0
=0,
\label{constraint_R}
\end{equation}
which imposes a constraint on the initial conditions. This relation is equivalent to the vanishing of the Hamiltonian density and encodes the residual reparametrization invariance inherited from the Polyakov formulation.

Substituting the constraint \eqref{constraint_R} into the expression for the canonical momentum \eqref{momentum_R}, the initial momentum is fixed as
\begin{equation}
P_R(\tau_0)
=
\pm
\frac{(t-\tau_0)^{\alpha-1}}{\Gamma(\alpha)}
\sqrt{1 - R_0^2\sin^2\theta_0},
\label{initialcond_R}
\end{equation}
with $R(\tau_0)=R_0$. The two branches correspond to initially expanding or contracting configurations of the loop, reflecting the two possible signs of the radial velocity compatible with the constraint.

As discussed earlier, the parameter $t$ represents the upper limit of the fractional integration and can be identified with the end of the radiation dominated era. In realistic cosmological settings this corresponds to a very large time scale, $t \simeq 1.6\times10^{12}\,\mathrm{sec}$. Consequently, there exists a large hierarchy between the initial time $\tau_0$ and the upper scale $t$. Direct numerical integration of Hamilton's equations \eqref{Rdot_eq} and \eqref{Pdot_eq} in terms of the physical time $\tau$ is therefore inconvenient, as it leads to poor resolution at early times and potential numerical instabilities over long evolution intervals.

To overcome this issue and achieve a more uniform resolution across many orders of magnitude in time, it is useful to introduce a logarithmic time variable defined by
\begin{equation}
y=\frac{1}{2}\log_{10}\left(\frac{\tau}{\tau_0}\right),
\label{y_def}
\end{equation}
so that $\tau=\tau_0\,10^{2y}$ and $d\tau/dy=2\ln(10)\,\tau$. This parametrization effectively stretches the early-time region while compressing late-time evolution, thereby improving numerical stability without altering the underlying dynamics. Expressing all dynamical quantities in terms of the variable $y$, Hamilton’s equations \eqref{Rdot_eq} and \eqref{Pdot_eq} can be rewritten as
\begin{eqnarray}
\frac{dR(y)}{dy}
&=&
\ln 10
\left[
R(y)+
\frac{2\Gamma(\alpha)\,\tau_0 10^{2y}}{(t-\tau_0 10^{2y})^{\alpha-1}}\,P_R(y)
\right],
\label{R_y_final}\\
\frac{dP_R(y)}{dy}
&=&
\ln 10
\left[
-P_R(y)
+
\frac{2\tau_0 10^{2y}(t-\tau_0 10^{2y})^{\alpha-1}}{\Gamma(\alpha)}\,R(y)\sin^2\theta_0
\right].
\label{P_y_final}
\end{eqnarray}

These equations define a closed dynamical system in terms of the logarithmic variable $y$, which is particularly well suited for numerical analysis over cosmological time scales. The appearance of the overall factor $\ln 10$ and the factor of $2$ inside the coefficients follows directly from the definition \eqref{y_def}, and reflects the specific choice of parametrization that optimally distributes the large time hierarchy between $\tau_0$ and $t$. The initial conditions are now naturally specified at $y=0$, corresponding to $\tau=\tau_0$, and take the form
\begin{equation}
R(0)=R_0, 
\qquad 
P_R(0)=
\pm
\frac{(t-\tau_0)^{\alpha-1}}{\Gamma(\alpha)}
\sqrt{1 - R_0^2\sin^2\theta_0},
\label{initialcond_y}
\end{equation}
where, once again, the choice of sign determines whether the loop is initially expanding or contracting. Moreover, one must have $R_0^2 \sin^2\theta_0 \leq 1$ for the initial momenta to be real. This initialization ensures consistency with the constraint structure of the theory and guarantees that the subsequent evolution governed by Eqs.~\eqref{R_y_final} and \eqref{P_y_final} remains confined to the physical phase space. For the numerical analysis we have considered initially expanding loop which corresponds to the positive sign of $P_R(0)$ in \eqref{initialcond_y}.

\begin{figure}[H]
    \centering
    \begin{minipage}{0.45\textwidth}
        \centering
        \includegraphics[width=7cm,height=6cm]{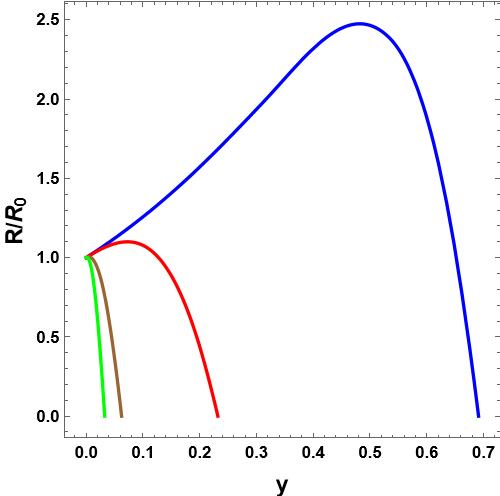}
    \end{minipage}
    \hspace{0.5cm}
    \begin{minipage}{0.45\textwidth}
        \centering
        \includegraphics[width=7cm,height=6cm]{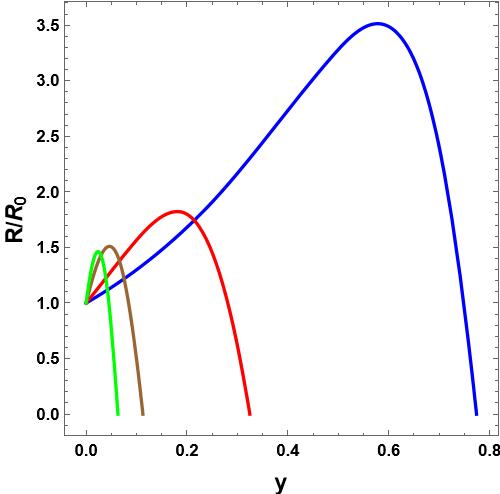}
    \end{minipage}

\caption{Evolution of the normalized loop radius $R/R_0$ as a function of $y$ for different initial times $\tau_0$. \textbf{Left panel:} $R_0=1$, $\theta_0=\pi/2$, $\alpha=0.75$. Curves correspond to $\tau_0=\{0.1,\,1,\,5,\,10\}$ sec (blue, red, brown, green), collapsing at $y=\{0.691226,\,0.231953,\,0.0623543,\,0.0325816\}$ respectively. \textbf{Right panel:} $R_0=1$, $\theta_0=\pi/4$, $\alpha=0.75$ with the same $\tau_0$ values. Collapse occurs at $y=\{0.774084,\,0.324634,\,0.112622,\,0.0631037\}$.} \label{fig:12}
\end{figure}

The numerical results displayed in Fig.~\ref{fig:12} reveal several important features of the dynamics of fractional cosmic string loops in an expanding background. First, for both panels, the loop radius $R/R_0$ initially increases for sufficiently small $\tau_0$, indicating that cosmological expansion dominates over string tension at early stages. However, this expansion is always transient, and the loop eventually reaches a maximum size before collapsing to $R=0$. This behavior reflects the competition between the Hubble stretching term and the effective tension term proportional to $\sin^2\theta_0$.

A clear trend emerges when varying the initial time $\tau_0$. As $\tau_0$ increases, the collapse occurs at progressively smaller values of the logarithmic time $y$. Physically, this indicates that loops formed later in the radiation dominated era have a shorter lifetime. This can be understood from the fact that at larger $\tau_0$, the Hubble rate $H=1/(2\tau)$ is smaller, reducing the expansion-driven stretching and allowing the tension to dominate more rapidly, leading to earlier collapse.

The comparison between the left and right panels highlights the role of the angular parameter $\theta_0$. For $\theta_0=\pi/4$, the collapse is systematically delayed compared to the case $\theta_0=\pi/2$. This behavior follows directly from the effective potential term $R^2\sin^2\theta_0$ in the Lagrangian. A smaller $\theta_0$ reduces the effective tension contribution, thereby allowing the loop to expand for a longer duration before collapsing. Another notable feature is that the initial growth phase becomes more pronounced for smaller $\tau_0$, particularly for the blue curves. This reflects the fact that at early cosmological times, the expansion term dominates strongly, leading to a significant temporary increase in the loop size before the eventual collapse.

We also observe that, for fixed values of $R_0$, $\tau_0$, and $\theta_0$, varying the fractional parameter $\alpha$ does not significantly affect the collapse time of the loop. This can be understood from the structure of the fractional kernel $(t-\tau)^{\alpha-1}$. For small $\tau_0$, the evolution occurs in a regime where $\tau \ll t$, with $t \sim 10^{12}\,\mathrm{sec}$. In this limit, the kernel becomes effectively constant, $(t-\tau)^{\alpha-1} \approx t^{\alpha-1}$, and thus factors out of the equations of motion without altering the qualitative dynamics. As a result, the loop evolution reduces effectively to the standard (non-fractional) case, explaining why the collapse time remains insensitive to $\alpha$ in this regime.

Overall, the plots demonstrate that the dynamics of fractional cosmic string loops is primarily governed by the interplay between cosmological expansion, encoded through $\tau_0$, and the effective tension controlled by $\theta_0$, while fractional effects become subdominant when the evolution occurs far from the memory scale set by $t$.

\subsection{Circular loop with time dependent polar angle}

We now consider a more general configuration in which the polar angle evolves dynamically along the worldsheet. In contrast to the previous case where the string loop was confined to a fixed plane, here the loop is allowed to move on the two-sphere of the FLRW spatial geometry. This introduces an additional dynamical degree of freedom associated with angular motion of the loop. The embedding of the string worldsheet is chosen as

\begin{equation}
\sigma^0=\tau,\qquad
\chi=\chi(\tau),\qquad
\theta=\theta(\tau),\qquad
\sigma^1=\phi .
\label{embedding2}
\end{equation}

In this configuration, both the comoving radial coordinate $\chi(\tau)$ and the polar angle $\theta(\tau)$ evolve with cosmic time. Physically, this describes a circular loop whose size changes while its orientation evolves on the two-sphere, thereby incorporating both radial contraction due to string tension and angular motion induced by the background geometry.

For the spatially flat universe ($K=0$), one has $\Sigma_K(\chi)=\chi$, and during the radiation dominated epoch the scale factor is given by Eq.~\eqref{radiationscale}. As in the previous subsection, it is convenient to introduce the physical radius $R(\tau)=a(\tau)\chi(\tau)$, which isolates the intrinsic dynamics of the loop from the cosmological expansion and allows a direct comparison with the fixed-angle configuration.

We again employ the fractional Polyakov action \eqref{FractionalAction}. The worldsheet metric can be written in the partially gauge-fixed form given in Eq.~\eqref{wmetric}. For the embedding \eqref{embedding2}, the induced metric remains diagonal since $\partial_\sigma \chi=\partial_\sigma \theta=0$, ensuring that the constraint $T_{\tau\sigma}=0$ is automatically satisfied. The spatial component of the induced metric depends only on $\tau$, and therefore the residual function $\omega(\tau,\sigma)$ can consistently be taken to depend only on $\tau$. Using the remaining Weyl symmetry, we fix $\omega(\tau,\sigma)=1$, so that the worldsheet metric reduces to the flat form $h_{ab}=\eta_{ab}$, exactly as in the fixed-angle case. Thus, the simplifications arising from symmetry persist even in the presence of angular dynamics.

Substituting the embedding \eqref{embedding2} into the Polyakov action and evaluating the kinetic and spatial contributions using the FLRW background, one finds that the Lagrangian acquires an additional contribution from the angular motion. Using the radiation-era relation $\dot{a}/a=1/(2\tau)$ and expressing the result in terms of the physical radius $R(\tau)$, the effective Lagrangian becomes

\begin{equation}
\mathcal{L}
=
-\frac{(t-\tau)^{\alpha-1}}{2\Gamma(\alpha)}
\left[
1
- \left(\dot{R}-\frac{R}{2\tau}\right)^2
- R^2 \dot{\theta}^2
- R^2 \sin^2\theta
\right].
\label{lagrangian_R_theta}
\end{equation}

Compared to the fixed-angle configuration, the dynamics now involves two coupled degrees of freedom. The term $R^2\dot{\theta}^2$ represents the kinetic energy associated with motion on the two-sphere, while $R^2\sin^2\theta$ plays the role of an effective potential. The combination $(\dot{R}-R/2\tau)$ retains its interpretation as the competition between intrinsic string dynamics and cosmological expansion, and the overall fractional factor $(t-\tau)^{\alpha-1}$ continues to encode nonlocal memory effects.

The Lagrangian \eqref{lagrangian_R_theta} defines a two-dimensional dynamical system with generalized coordinates $R(\tau)$ and $\theta(\tau)$. The canonical momenta are obtained by differentiating with respect to the corresponding velocities. The momentum conjugate to $R(\tau)$ retains the same structure as in the fixed-angle case and is given by
\begin{equation}
P_R(\tau)
=
\frac{(t-\tau)^{\alpha-1}}{\Gamma(\alpha)}
\left(\dot{R}-\frac{R}{2\tau}\right),
\label{PR_theta}
\end{equation}
while the momentum conjugate to $\theta(\tau)$ is
\begin{equation}
P_\theta(\tau)
=
\frac{(t-\tau)^{\alpha-1}}{\Gamma(\alpha)}\,R^2 \dot{\theta}.
\label{Ptheta}
\end{equation}

The quantity $P_\theta(\tau)$ can be interpreted as an effective angular momentum. However, unlike the standard string case, it is not conserved due to the explicit time dependence introduced by both the cosmological background and the fractional kernel.

The Hamiltonian is obtained via the Legendre transformation $\mathcal{H}=P_R\dot{R}+P_\theta\dot{\theta}-\mathcal{L}$, where the velocities are expressed in terms of the canonical variables. Using Eqs.~\eqref{PR_theta} and \eqref{Ptheta}, one finds

\begin{equation}
\mathcal{H}
=
\frac{\Gamma(\alpha)}{2\,(t-\tau)^{\alpha-1}}
\left(
P_R^2+\frac{P_\theta^2}{R^2}
\right)
+
\frac{P_R R}{2\tau}
+
\frac{(t-\tau)^{\alpha-1}}{2\Gamma(\alpha)}
\left[
1 - R^2 \sin^2\theta
\right].
\label{Hamiltonian_theta}
\end{equation}

The structure of the Hamiltonian closely parallels that of the fixed-angle case, with the addition of a centrifugal term $P_\theta^2/R^2$. The first term represents the kinetic energy in phase space, the second term arises purely from cosmological expansion and couples $R$ and $P_R$, while the last term acts as an effective potential. The explicit time dependence signals that the system is non-autonomous and reflects both expansion and memory effects. The equations of motion follow from Hamilton's equations and can be given as 
\begin{eqnarray}
\dot{R}
&=&
\frac{\Gamma(\alpha)}{(t-\tau)^{\alpha-1}}\,P_R
+
\frac{R}{2\tau},
\label{Rdot_theta}\\
\dot{\theta}
&=&
\frac{\Gamma(\alpha)}{(t-\tau)^{\alpha-1}}\,
\frac{P_\theta}{R^2},
\label{thetadot}\\
\dot{P}_R
&=&
-\frac{P_R}{2\tau}
+
\frac{\Gamma(\alpha)}{(t-\tau)^{\alpha-1}}\,
\frac{P_\theta^2}{R^3}
+
\frac{(t-\tau)^{\alpha-1}}{\Gamma(\alpha)}\,R \sin^2\theta,
\label{PRdot_theta}\\
\dot{P}_\theta
&=&
-\frac{(t-\tau)^{\alpha-1}}{\Gamma(\alpha)}\,R^2 \sin\theta \cos\theta.
\label{Pthetadot}
\end{eqnarray}

These equations form a coupled dynamical system describing the joint radial and angular evolution of the loop. The term proportional to $P_\theta^2/R^3$ represents the centrifugal contribution arising from angular motion, while the angular equation shows that the effective angular momentum evolves under a potential proportional to $\sin^2\theta$. The terms proportional to $1/\tau$ again encode the effect of cosmological expansion and act as damping contributions. The fractional parameter $\alpha$ enters through time-dependent coefficients, modifying both inertial and potential terms and introducing memory effects.

As in the fixed-angle case, consistency of the dynamics requires the imposition of the worldsheet constraint $T_{\tau\tau}=0$, which now takes the form
\begin{equation}
1
-
\left(\dot{R}-\frac{R}{2\tau}\right)^2
-
R^2 \dot{\theta}^2
-
R^2 \sin^2\theta
=0.
\label{constraint_theta}
\end{equation}

This constraint defines the physical phase space and must be satisfied by the initial data at $\tau=\tau_0$. Since the system involves two dynamical variables, $R(\tau)$ and $\theta(\tau)$, the constraint removes one degree of freedom, ensuring that the evolution remains consistent with the residual reparametrization invariance of the Polyakov formulation. A natural and physically motivated choice is to consider configurations in which the loop is initially comoving with the cosmological expansion, namely
\begin{equation}
\left(\dot{R}-\frac{R}{2\tau}\right)_{\tau=\tau_0}=0.
\label{IC_case1}
\end{equation}
In this case the radial momentum vanishes, $P_R(\tau_0)=0$, and the constraint fixes the angular velocity, leading to
\begin{equation}
P_\theta(\tau_0)
=
\pm
\frac{(t-\tau_0)^{\alpha-1}}{\Gamma(\alpha)}\,R_0^2
\sqrt{\frac{1}{R_0^2}-\sin^2\theta_0}.
\label{Ptheta_case1}
\end{equation}

Thus, once $R_0$ and $\theta_0$ are specified, the angular dynamics is completely determined by the constraint. Alternatively, one may consider configurations with vanishing initial angular motion,
\begin{equation}
P_\theta(\tau_0)=0,
\qquad
\theta(\tau_0)=\theta_0,
\label{IC_case2}
\end{equation}
for which the constraint reduces to the fixed-angle condition and determines the initial radial momentum as
\begin{equation}
P_R(\tau_0)
=
\pm
\frac{(t-\tau_0)^{\alpha-1}}{\Gamma(\alpha)}
\sqrt{1 - R_0^2 \sin^2\theta_0}.
\label{PR_case2}
\end{equation}

Given the Hamiltonian equations of motion and the initial conditions specified above, the next step is to investigate the dynamical evolution of the system numerically. To ensure numerical stability across the wide hierarchy of cosmological time scales, it is convenient to reformulate the system in terms of the logarithmic time variable introduced earlier, cf.\ Eq.~\eqref{y_def}. This reparametrization enables a uniform resolution of both early- and late-time dynamics without altering the physical content of the evolution during the radiation-dominated era. In terms of the variable $y$, the coupled system \eqref{Rdot_theta}--\eqref{Pthetadot} can be written as
\begin{eqnarray}
\frac{dR}{dy}
&=&
\ln 10\left[
R
+
\frac{2\Gamma(\alpha)\,\tau_0 10^{2y}}{(t-\tau_0 10^{2y})^{\alpha-1}}\,P_R
\right],
\label{R_y_theta}\\
\frac{d\theta}{dy}
&=&
\ln 10\left[
\frac{2\Gamma(\alpha)\,\tau_0 10^{2y}}{(t-\tau_0 10^{2y})^{\alpha-1}}
\frac{P_\theta}{R^2}
\right],
\label{theta_y}\\
\frac{dP_R}{dy}
&=&
\ln 10\left[
- P_R
+
\frac{2\Gamma(\alpha)\,\tau_0 10^{2y}}{(t-\tau_0 10^{2y})^{\alpha-1}}
\frac{P_\theta^2}{R^3}
+
\frac{2\tau_0 10^{2y}(t-\tau_0 10^{2y})^{\alpha-1}}{\Gamma(\alpha)}
R\sin^2\theta
\right],
\label{PR_y_theta}\\
\frac{dP_\theta}{dy}
&=&
\ln 10\left[
-
\frac{2\tau_0 10^{2y}(t-\tau_0 10^{2y})^{\alpha-1}}{\Gamma(\alpha)}
R^2 \sin\theta\cos\theta
\right].
\label{Ptheta_y}
\end{eqnarray}

These equations provide a numerically efficient representation of the dynamics. The logarithmic time variable effectively rescales the evolution, allowing for controlled integration over both early and late cosmological epochs. The initial conditions are imposed at $y=0$, corresponding to $\tau=\tau_0$, and are specified as
\begin{equation}
R(0)=R_0, 
\qquad 
\theta(0)=\theta_0,
\end{equation}
whereas the remaining initial data are determined by the constraint \eqref{constraint_theta}, leading to two distinct classes of physically relevant configurations.

\vspace{0.3cm}
\noindent
\textbf{(i) Angularly driven evolution (initially comoving configuration):}
\begin{equation}
P_R(0)=0, 
\qquad
P_\theta(0)=
\pm
\frac{(t-\tau_0)^{\alpha-1}}{\Gamma(\alpha)}\,
\sqrt{1-R_0^2\sin^2\theta_0}.\label{InitialCondAngular}
\end{equation}

In this configuration, the condition $P_R(0)=0$ implies
$\dot{R} - \frac{R}{2\tau} = 0$ at $\tau=\tau_0$, meaning that the loop is initially comoving with the Hubble expansion. Consequently, there is no intrinsic radial motion at the initial time, and the subsequent evolution of $R(\tau)$ arises from the interplay between cosmological expansion and string tension. The angular momentum $P_\theta(0)$ is not a free parameter but is completely fixed by the constraint \eqref{constraint_theta}. This ensures that the angular motion is dynamically induced rather than externally prescribed. The requirement that $P_\theta(0)$ be real imposes the condition
\begin{equation}
R_0^2 \sin^2\theta_0 \leq 1,
\end{equation}
which restricts the allowed initial configurations. Physically, this case describes loops whose initial dynamics is dominated by angular motion on the two-sphere, with centrifugal effects emerging naturally from the constraint.

\vspace{0.3cm}
\noindent
\textbf{(ii) Purely radial evolution (vanishing initial angular momentum):}
\begin{equation}
P_\theta(0)=0,
\qquad
P_R(0)=
\pm
\frac{(t-\tau_0)^{\alpha-1}}{\Gamma(\alpha)}
\sqrt{1-R_0^2\sin^2\theta_0}.\label{InitialCondRadial}
\end{equation}

Here, the absence of angular momentum implies $\dot{\theta}(0)=0$, so the loop does not initially move on the two-sphere and its orientation remains momentarily fixed. The entire dynamics is therefore governed by the radial degree of freedom $R(\tau)$.The constraint \eqref{constraint_theta} determines the initial radial momentum $P_R(0)$, which encodes the competition between string tension and the effective potential term $R^2\sin^2\theta_0$. The reality of $P_R(0)$ requires
\begin{equation}
R_0^2 \sin^2\theta_0 \leq 1,
\end{equation}
which defines the physically admissible region of initial data. Depending on the sign choice, the loop may initially expand or contract, leading to qualitatively different subsequent evolutions. In this case, any angular motion that develops at later times is generated dynamically through the coupling in the equations of motion, rather than being present in the initial state.

\begin{figure}[H]
\centering
\begin{minipage}{0.45\textwidth}
    \centering
    \includegraphics[width=7cm,height=6cm]{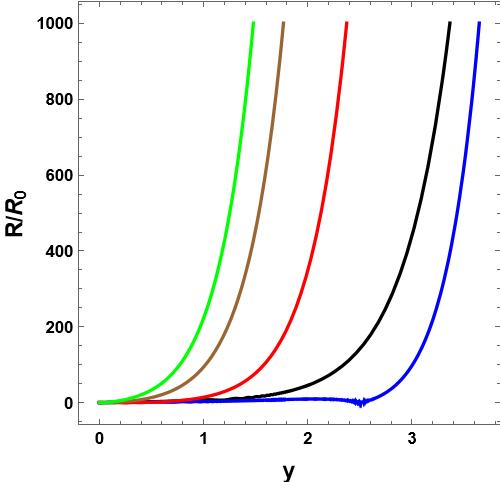}
    \vspace{-0.3cm}
\end{minipage}
\hfill
\begin{minipage}{0.45\textwidth}
    \centering
    \includegraphics[width=7cm,height=6cm]{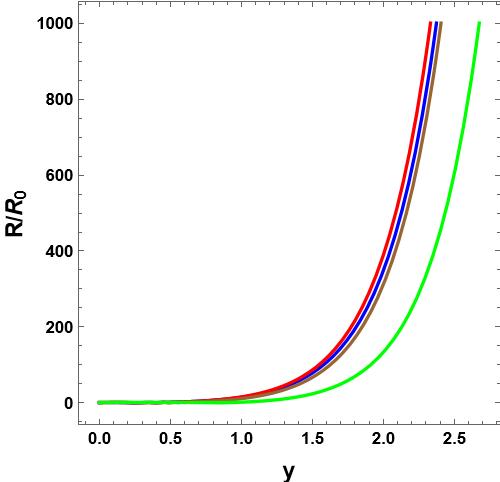}
    \vspace{-0.3cm}
\end{minipage}
\caption{Evolution of the normalized physical radius $R/R_0$ as a function of the logarithmic time variable $y$ for angularly driven initial conditions. \textbf{Left panel:} Dependence on the fractional time parameter $t$ for fixed $R_0=1$, $\alpha=0.75$, and $\theta_0=\pi/3$. The curves correspond to $t=\{0.01$ (black), $0.1$ (blue), $1$ (red), $5$ (brown), $10$ (green)$\}$.\textbf{Right panel:} Dependence on the initial polar angle $\theta_0$ for fixed $R_0=1$, $\tau_0=1$. The curves correspond to $\theta_0=\{\pi/3$ (blue), $\pi/4$ (red), $\pi/6$ (brown), $\pi/12$ (green)$\}$.
}
\label{fig:angular_case}
\end{figure}

The evolution corresponding to the angularly driven initial condition \eqref{InitialCondAngular} is illustrated in Fig.~\ref{fig:angular_case}. In this setup, the system is initialized with vanishing radial momentum, $P_R(0)=0$, while the angular momentum is entirely determined by the constraint. Consequently, the subsequent dynamics is intrinsically two-dimensional and governed by the coupled interplay between cosmological expansion, fractional memory effects, and constraint-induced angular motion.

The left panel demonstrates the dependence on the fractional time parameter $t$. Increasing $t$ systematically delays the onset of rapid growth of $R(\tau)$, indicating that the fractional kernel $(t-\tau)^{\alpha-1}$ suppresses the early-time dynamics and effectively postpones the transfer of energy into radial expansion. For smaller values of $t$, this suppression is weaker, and the loop enters the accelerated expansion phase earlier. This behavior highlights a key feature of the fractional framework: the temporal nonlocality directly controls the onset of dynamical instability.

The right panel shows the dependence on the initial polar angle $\theta_0$, which fixes the initial angular momentum through the constraint. Larger $\theta_0$ enhances $P_\theta(0)$ and therefore strengthens the effective centrifugal contribution. As a result, the system departs earlier from the quasi-comoving regime and undergoes a more rapid expansion. In contrast, smaller $\theta_0$ suppresses the angular contribution, leading to a prolonged quasi-static phase in which the loop closely follows the background expansion before transitioning to accelerated growth. 

A central observation is that the angular degree of freedom acts as an effective driving mechanism: once the centrifugal term becomes dominant, it triggers a rapid growth of the loop radius. This provides a qualitatively new channel for expansion that is absent in fixed-angle configurations.

\begin{figure}[H]
\centering
\begin{minipage}{0.45\textwidth}
    \centering
    \includegraphics[width=7cm,height=6cm]{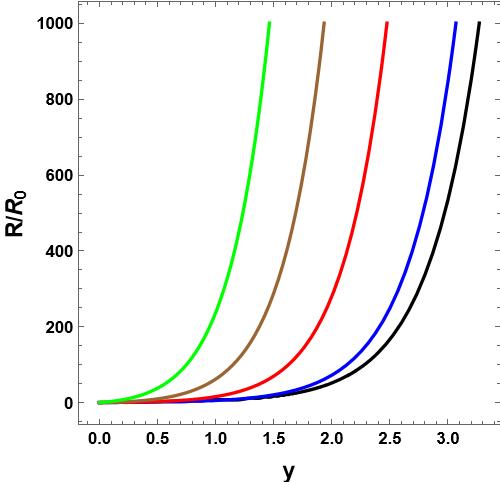}
    \vspace{-0.3cm}
\end{minipage}
\hfill
\begin{minipage}{0.45\textwidth}
    \centering
    \includegraphics[width=7cm,height=6cm]{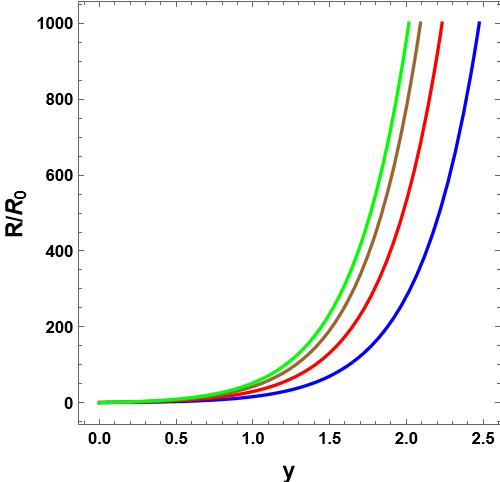}
    \vspace{-0.3cm}
\end{minipage}
\caption{Evolution of the normalized physical radius $R/R_0$ as a function of the logarithmic time variable $y$ for purely radial evolution initial conditions. \textbf{Left panel:} Dependence on the fractional time parameter $t$ for fixed $R_0=1$, $\alpha=0.75$, and $\theta_0=\pi/3$. The curves correspond to $t=\{0.01$ (black), $0.1$ (blue), $1$ (red), $5$ (brown), $10$ (green)$\}$.\textbf{Right panel:} Dependence on the initial polar angle $\theta_0$ for fixed $R_0=1$, $\tau_0=1$. The curves correspond to $\theta_0=\{\pi/3$ (blue), $\pi/4$ (red), $\pi/6$ (brown), $\pi/12$ (green)$\}$.
}
\label{fig:radial_case}
\end{figure}

The corresponding evolution for purely radial initial conditions \eqref{InitialCondRadial} is presented in Fig.~\ref{fig:radial_case}. In this case, the system is initialized with vanishing angular momentum, $P_\theta(0)=0$, so that $\dot{\theta}(0)=0$ and the loop begins without motion on the two-sphere. The early-time dynamics is therefore purely radial, with the initial radial momentum fixed by the constraint.

The left panel again shows that increasing $t$ delays the onset of rapid growth, confirming that the suppressive effect of the fractional kernel is robust and independent of the initial angular configuration. However, in contrast to the angularly driven case, the absence of initial angular momentum implies that the expansion is not immediately supported by a centrifugal mechanism.

The right panel illustrates that the initial angle $\theta_0$ still plays a nontrivial role through the constraint, which determines $P_R(0)$. Larger $\theta_0$ increases the effective potential contribution $R^2\sin^2\theta_0$, thereby enhancing the initial radial momentum and leading to an earlier onset of expansion. Smaller $\theta_0$ results in a weaker effective driving and a delayed transition to the growth phase.

A key qualitative difference between the two cases is that, in the purely radial setup, the expansion is initially governed solely by the competition between string tension and cosmological expansion, with angular effects emerging only dynamically at later times. In contrast, for angularly driven initial conditions, the centrifugal contribution is present from the outset and significantly accelerates the growth of the loop.

An important and novel outcome of our analysis is the emergence of expanding loop solutions in a regime where conventional treatments typically predict inevitable collapse. In particular, for $\theta(0)=\pi/2$, the dynamics generically leads to collapsing configurations for all values of $R_0$ and $\tau_0$, consistent with the constraint $R_0^2 \sin^2\theta_0 \leq 1$, and this behavior persists for both angularly driven and purely radial initial conditions.

In contrast, once the polar angle $\theta(\tau)$ is promoted to a dynamical variable, the qualitative behavior of the system changes significantly. We find that the coupled evolution of $R(\tau)$ and $\theta(\tau)$ admits a new class of solutions in which the loop undergoes sustained and accelerated expansion. This transition is driven by the dynamical buildup of angular motion, which generates an effective centrifugal contribution that can counteract the contracting influence of string tension.

This result demonstrates that allowing angular dynamics is not merely a quantitative extension of the standard setup, but leads to a qualitative modification of the phase space structure and the late-time fate of the loop. In particular, the combined effect of angular motion and fractional memory introduces a mechanism that can prevent collapse and instead drive expansion. Such expanding solutions are absent in fixed-angle analyses and therefore represent a genuinely new feature of the present framework.

\subsection{Emergence of chaos in a circular loop with time dependent polar angle}

The coupled system \eqref{R_y_theta}--\eqref{Ptheta_y} describes a nonlinear, non-autonomous dynamical system with two degrees of freedom. The presence of explicit time dependence through the fractional kernel, together with nonlinear coupling between the radial and angular sectors, suggests that the dynamics may exhibit complex behavior beyond simple monotonic expansion or collapse. In particular, the competition between the effective centrifugal term, the angular potential, and the time-dependent driving induced by the fractional kernel provides a natural setting for the emergence of chaotic dynamics. To explore this possibility, it is useful to first analyze the structure of equilibrium configurations and their stability properties.

\subsubsection{Fixed point analysis}

Before addressing the onset of chaotic behavior, it is useful to analyze the fixed-point structure of the dynamical system \eqref{R_y_theta}--\eqref{Ptheta_y}. Since the system depends explicitly on the logarithmic time variable $y$, it is inherently non-autonomous and does not admit fixed points in the strict sense. Nevertheless, one can still extract meaningful information by considering instantaneous (or frozen-time) fixed points, obtained by setting the right-hand sides of the evolution equations to zero at a given value of $y$.

For notational convenience, we define the time-dependent coefficients
\begin{equation}
A(y)=\frac{2\Gamma(\alpha)\,\tau_0 10^{2y}}{(t-\tau_0 10^{2y})^{\alpha-1}},
\qquad
B(y)=\frac{2\tau_0 10^{2y}(t-\tau_0 10^{2y})^{\alpha-1}}{\Gamma(\alpha)}.
\end{equation}
The fixed-point conditions can then be expressed compactly as
\begin{equation}
R + A(y) P_R = 0, \qquad \frac{A(y)}{R^2} P_\theta = 0,
\end{equation}
together with
\begin{equation}
- P_R + A(y)\frac{P_\theta^2}{R^3} + B(y) R \sin^2\theta = 0,
\qquad
- B(y) R^2 \sin\theta \cos\theta = 0.
\end{equation}

We restrict attention to finite configurations with $R \neq 0$. The second condition then immediately implies $P_\theta=0$. Substituting this into the angular equation yields $\sin\theta \cos\theta = 0$. Since the physically relevant domain is the principal branch $\theta \in [0,\pi/2]$, the only admissible solutions are $\theta_*=0$ and $\theta_*=\pi/2$.

The remaining conditions determine the radial momentum as $P_R = -R/A(y)$, together with the consistency requirement
\begin{equation}
\frac{1}{A(y)} + B(y)\sin^2\theta = 0.
\end{equation}
Because both $A(y)$ and $B(y)$ are strictly positive in the physical regime, this condition can only be satisfied if $\sin^2\theta=0$, which singles out $\theta_*=0$ as the only admissible instantaneous fixed point. Importantly, no consistent solution exists at $\theta=\pi/2$, indicating that this configuration cannot support equilibrium.

To understand the local behavior, we linearize the angular dynamics by writing $\theta=\theta_*+\delta\theta$. Near $\theta_*=0$, one finds that the leading contribution behaves as $dP_\theta/dy \sim -B(y) R^2 \delta\theta$, which corresponds to a restoring force and thus local stability in the angular direction. In contrast, near $\theta=\pi/2$, the linear term vanishes identically, and the evolution is governed by higher-order nonlinear terms, signaling marginal or unstable behavior.

This structure has important dynamical consequences. The absence of a consistent fixed point at $\theta=\pi/2$, together with the explicit time dependence and strong nonlinear coupling between $R$ and $\theta$, implies that the system does not possess a global equilibrium configuration. Instead, trajectories evolve in a time-dependent phase space where the effective forces continuously change due to both the fractional kernel and angular backreaction.

Such a setting naturally favors complex dynamics. The competition between the centrifugal contribution and the effective angular potential introduces a time-dependent nonlinear driving that can repeatedly stretch and redirect trajectories in phase space. Combined with the non-autonomous nature of the system, this provides the necessary ingredients for the emergence of chaos, characterized by sensitive dependence on initial conditions and irregular evolution.

These observations indicate that promoting $\theta(\tau)$ to a dynamical variable fundamentally alters the phase-space structure of the system. In particular, the interplay of angular dynamics and fractional memory effects not only destabilizes certain configurations but also opens the possibility of chaotic behavior in the evolution of circular cosmic string loops.

\subsubsection{Lyapunov exponent and diagnostics of chaos}

To quantitatively characterize the presence of chaos in the dynamical system \eqref{R_y_theta}--\eqref{Ptheta_y}, we compute the Lyapunov spectrum for the angularly driven evolution (initially comoving configuration), which measures the exponential rates at which nearby trajectories in phase space diverge or converge. A positive Lyapunov exponent is the hallmark of sensitive dependence on initial conditions and therefore signals chaotic dynamics \cite{Benettin1980,Wolf1985,Ott2002}. Let ${\bf X}(y) = (R,\theta,P_R,P_\theta)$ denote the phase-space vector. The evolution of an infinitesimal perturbation $\delta{\bf X}(y)$ around a reference trajectory is governed by the linearized (variational) equations
\begin{equation}
\frac{d}{dy}\,\delta{\bf X}(y) = \mathbf{J}(y)\,\delta{\bf X}(y),
\end{equation}
where $\mathbf{J}(y) = \partial \dot{\bf X}/\partial {\bf X}$ is the Jacobian matrix evaluated along the trajectory. This relation describes how small perturbations evolve under the local flow of the system.

To extract the full Lyapunov spectrum, we evolve the background trajectory together with a set of four linearly independent deviation vectors spanning the tangent space. Since these vectors generically grow (or shrink) exponentially and tend to align along the most unstable direction, we periodically re-orthonormalize them using a Gram--Schmidt procedure. At each re-orthonormalization step, the logarithmic growth factors of these vectors are recorded. The Lyapunov exponents are then obtained as time-averaged growth rates \cite{Sandri1996},
\begin{equation}
\lambda_i = \lim_{y\to\infty} \frac{1}{y} \sum_{k} \ln \frac{\|\delta{\bf X}_i^{(k)}\|}{\|\delta{\bf X}_i^{(k-1)}\|}.
\label{lyapunov_def2}
\end{equation}

Since the phase space of the system is four-dimensional, this procedure yields a set of four Lyapunov exponents,
\begin{equation}
(\lambda_R,\lambda_\theta,\lambda_{P_R},\lambda_{P_\theta}),
\label{lyapunov_spec2}
\end{equation}
which characterize the stability properties along different phase-space directions. More specifically, $\lambda_R$ quantifies the rate of separation of nearby trajectories in the radial direction $R$, capturing radial instability, while $\lambda_\theta$ measures the sensitivity of the angular motion. The exponents $\lambda_{P_R}$ and $\lambda_{P_\theta}$ encode the stability properties associated with the conjugate momenta. In this way, the spectrum provides a directional decomposition of instability rather than a single averaged measure. The structure of the spectrum further reflects the underlying phase-space dynamics. In particular, the appearance of pairs of exponents with opposite signs indicates that local expansion along certain directions is accompanied by contraction along others. This behavior is characteristic of constrained dynamical systems, where stretching and folding mechanisms govern the evolution and ultimately lead to chaotic dynamics.

The present system naturally incorporates several ingredients conducive to chaos. The explicit $y$-dependence induced by the fractional kernel renders the system non-autonomous, while nonlinear couplings between $R$ and $\theta$ introduce competing dynamical effects. In particular, the centrifugal term $\sim P_\theta^2/R^3$ amplifies local instabilities, whereas the angular interaction $\sim R^2\sin\theta\cos\theta$ generates nonlinear feedback. Together, these mechanisms lead to repeated stretching and folding of phase-space trajectories, which is the defining signature of chaos. 

\begin{table}[h]
\centering
\begin{tabular}{|c|c|c|c|c|}
\hline
$\theta_0$ & $\lambda_R$ & $\lambda_\theta$ & $\lambda_{P_R}$ & $\lambda_{P_\theta}$ \\ \hline
$\pi/3$ & 2.30259 & 1.42345 & -2.30259 & -1.42345 \\ \hline
$\pi/4$ & 2.30259 & $2.58\times 10^{-12}$ & -2.30259 & $-2.57\times 10^{-12}$ \\ \hline
$\pi/6$ & 2.30259 & 1.42345 & -2.30259 & -1.42345 \\ \hline
$\pi/10$ & 2.30259 & 1.45766 & -2.30259 & -1.45766 \\ \hline
\end{tabular}
\caption{Lyapunov exponents for varying initial angle $\theta_0$ with $\alpha=0.75$, $\tau_0=10^{-16}$, and $R_0=1$.}
\label{tab:theta_variation}
\end{table}

We first fix $\alpha=0.75$, $\tau_0=10^{-16}$, and $R_0=1$, and vary the initial angle $\theta_0$. The results are presented in Table~\ref{tab:theta_variation}. A notable feature is that the radial exponent $\lambda_R$ remains positive and nearly constant across all configurations, indicating robust and persistent exponential instability in the radial direction. In contrast, $\lambda_\theta$ exhibits a strong dependence on $\theta_0$ and can become nearly vanishing (e.g., for $\theta_0=\pi/4$), signaling a temporary suppression of angular instability and the emergence of marginally stable angular motion.

\begin{table}[H]
\centering
\begin{tabular}{|c|c|c|c|c|}
\hline
$R_0$ & $\lambda_R$ & $\lambda_\theta$ & $\lambda_{P_R}$ & $\lambda_{P_\theta}$ \\ \hline
1/2 & 2.30259 & 1.3249 & -2.30259 & -1.3249 \\ \hline
1/4 & 2.30259 & 1.22635 & -2.30259 & -1.22635 \\ \hline
1/10 & 2.30259 & 1.09607 & -2.30259 & -1.09607 \\ \hline
1/25 & 2.30259 & 0.965792 & -2.30259 & -0.965792 \\ \hline
\end{tabular}
\caption{Lyapunov exponents for varying initial radius $R_0$ with $\alpha=0.75$, $\tau_0=10^{-16}$, and $\theta_0=\pi/3$.}
\label{tab:R_variation}
\end{table}

Next, fixing $\alpha=0.75$, $\tau_0=10^{-16}$, and $\theta_0=\pi/3$, we vary the initial radius $R_0$. As shown in Table~\ref{tab:R_variation}, the radial exponent $\lambda_R$ again remains essentially unchanged, demonstrating that radial chaos is largely insensitive to the initial size of the loop. In contrast, $\lambda_\theta$ decreases monotonically with decreasing $R_0$, indicating that smaller loops experience weaker angular instability. This points to a scale-dependent suppression of angular chaos.

\begin{table}[H]
\centering
\begin{tabular}{|c|c|c|c|c|}
\hline
$\alpha$ & $\lambda_R$ & $\lambda_\theta$ & $\lambda_{P_R}$ & $\lambda_{P_\theta}$ \\ \hline
0.95 & 2.30259 & 1.83523 & -2.30259 & -1.83523 \\ \hline
0.5 & 2.30259 & 0.897788 & -2.30259 & -0.897788 \\ \hline
0.2 & 2.30259 & 0.23088 & -2.30259 & -0.23088 \\ \hline
0.05 & 2.30259 & $2.84\times 10^{-4}$ & -2.30259 & $-2.84\times 10^{-4}$ \\ \hline
\end{tabular}
\caption{Lyapunov exponents for varying fractional parameter $\alpha$ with $R_0=1$, $\theta_0=\pi/6$, and $\tau_0=10^{-16}$.}
\label{tab:alpha_variation}
\end{table}

We now examine the role of fractional memory by varying $\alpha$ while keeping $R_0=1$, $\theta_0=\pi/6$, and $\tau_0=10^{-16}$ fixed. The results in Table~\ref{tab:alpha_variation} show that $\lambda_\theta$ decreases significantly as $\alpha$ is reduced and approaches zero in the small-$\alpha$ limit. This demonstrates that fractional memory effects act as an effective stabilizing mechanism for angular motion, progressively suppressing chaotic behavior in that sector. Remarkably, the radial exponent $\lambda_R$ remains largely unaffected, indicating a clear decoupling between radial and angular instability channels.
\begin{figure}[H]
\centering
\begin{minipage}{0.45\textwidth}
\centering
\includegraphics[width=7cm,height=6cm]{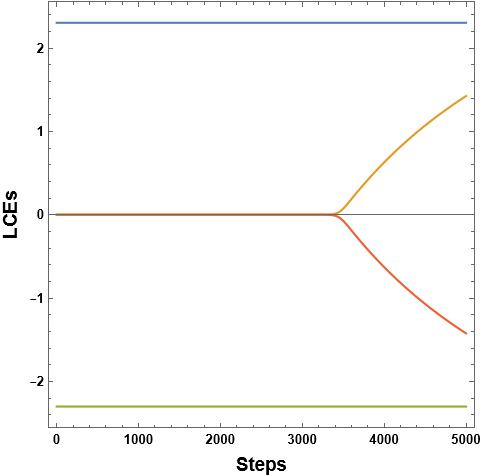}
\end{minipage}
\hfill
\begin{minipage}{0.45\textwidth}
\centering
\includegraphics[width=7cm,height=6cm]{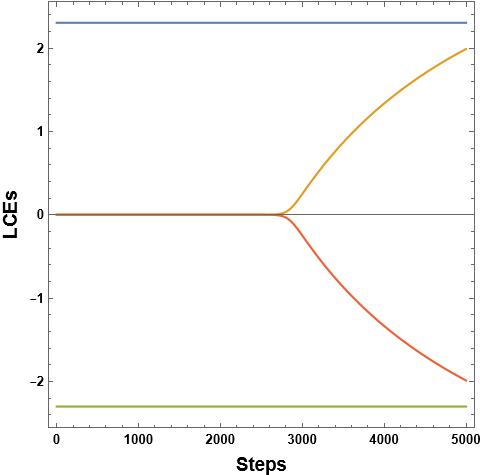}
\end{minipage}
\vspace{0.3cm}
\begin{minipage}{0.45\textwidth}
\centering
\includegraphics[width=7cm,height=6cm]{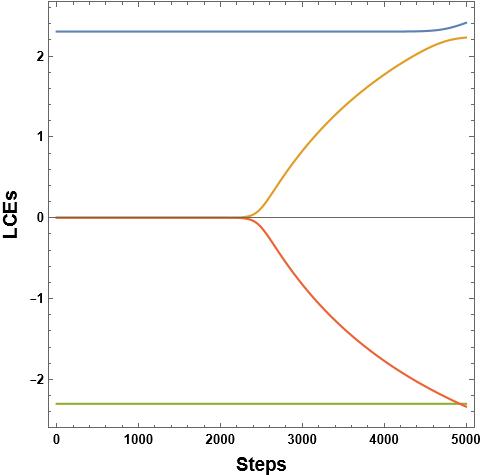}
\end{minipage}
\hfill
\begin{minipage}{0.45\textwidth}
\centering
\includegraphics[width=7cm,height=6cm]{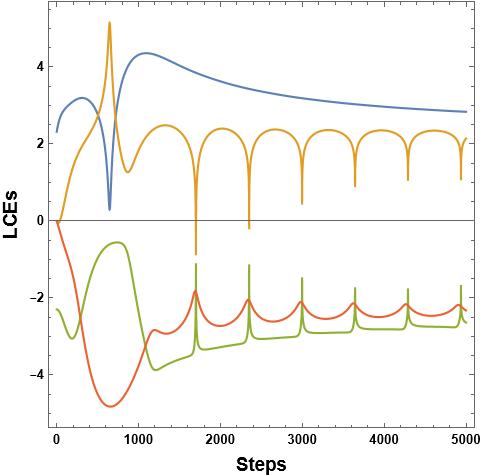}
\end{minipage}
\caption{Evolution of the Lyapunov spectrum as a function of integration steps for fixed $R_0=1$, $\alpha=0.75$, and $\theta_0=\pi/3$. The panels correspond to $\tau_0=10^{-16}, 10^{-8}, 10^{-5}, 1$ (top left to bottom right).}
\label{fig:lyapunov_spectrum}
\end{figure}

The convergence of the Lyapunov spectrum is illustrated in Fig.~\ref{fig:lyapunov_spectrum}, where the exponents are plotted as functions of the number of integration steps for different values of $\tau_0$. For very small $\tau_0$, corresponding to early formation times, the largest Lyapunov exponent rapidly stabilizes at a positive value with pronounced fluctuations, clearly indicating strongly chaotic dynamics. As $\tau_0$ is increased, a qualitative change in the behavior is observed. The fluctuations in the spectrum are significantly reduced, and the largest exponent approaches smaller values, signaling a weakening of chaos. For sufficiently large $\tau_0$, the system exhibits a transition toward marginal or weakly chaotic dynamics, with the Lyapunov exponents tending toward near-zero values. This demonstrates that the initial formation time of the loop plays a crucial role in controlling the degree of chaos. A similar qualitative behavior is expected for initial conditions corresponding to purely radial evolution (vanishing initial angular momentum), indicating that the suppression of chaos at late formation times is a robust feature of the dynamics rather than a consequence of angular motion alone.

A central outcome of this analysis can be summarized as follows: \emph{the system exhibits robust radial chaos across all parameter regimes, while angular instability is tunable and can be strongly suppressed by fractional memory effects and initial conditions}. Furthermore, the transition from strongly chaotic to weakly chaotic or near-regular dynamics is governed by the initial time $\tau_0$, providing a new control parameter for the dynamical behavior. Importantly, the presence of chaos is not merely incidental but is directly tied to the emergence of expanding loop solutions. The amplification of angular motion, captured by positive $\lambda_\theta$, enables an effective transfer of energy into the radial sector through the centrifugal term. This mechanism allows trajectories to overcome collapse and enter a regime of sustained expansion. Thus, chaotic dynamics, angular degrees of freedom, and fractional memory effects together provide the dynamical origin of the expanding solutions identified in this work.

\section{Conclusions and Discussion}

In this work we have developed a consistent framework to study the evolution of circular cosmic string loops in an expanding FLRW universe within a fractional Polyakov formulation. The key novelty of the approach lies in the introduction of a fractional action-like variational principle, which naturally incorporates nonlocal memory effects into the worldsheet dynamics. As a consequence, the resulting system is intrinsically non-autonomous and exhibits features that are absent in the standard Nambu--Goto description.

A central outcome of our analysis is that the dynamics of cosmic string loops is significantly enriched once both fractional effects and angular degrees of freedom are taken into account. In the fixed-angle configuration, the evolution is governed by the familiar competition between string tension and cosmological expansion. In this case, we recover the standard qualitative behavior: loops undergo a transient expansion phase but eventually collapse, with the collapse time controlled by the initial formation time and the effective tension parameter. This is consistent with earlier studies of loop evolution in expanding backgrounds, where collapse is typically unavoidable unless special conditions are imposed.

However, once the polar angle is promoted to a dynamical variable, the qualitative structure of the dynamics changes in a fundamental way. The system becomes genuinely two-dimensional in phase space, with a nontrivial coupling between radial and angular degrees of freedom. The angular motion generates an effective centrifugal contribution, which competes with the contracting influence of string tension. \textit{A key new result of this work is that this interplay gives rise to expanding loop solutions, which are absent in conventional treatments.} In particular, previous analyses in de Sitter or FLRW backgrounds have shown that loops generally collapse unless they are sufficiently large or satisfy restrictive conditions on their tension \cite{Larsen1994}, or unless additional assumptions such as time-dependent tension are introduced, which still impose nontrivial constraints on the dynamics as discussed in \cite{Yamaguchi2006}. In contrast, our results demonstrate that \textit{dynamical angular motion alone provides a new mechanism that can sustain expansion}, without requiring fine-tuning of the initial size or tension profile.

An important observation emerging from our study is the special role played by the configuration $\theta=\pi/2$. Even in the fully dynamical setup where $\theta(\tau)$ evolves with time, this configuration corresponds to a regime in which the effective potential is maximized and the centrifugal support is insufficient to counterbalance the string tension. \textit{As a result, trajectories initialized near $\theta=\pi/2$ generically lead to collapse, irrespective of the presence of angular dynamics.} This indicates that the expanding solutions arise only within a restricted region of phase space away from this maximally stretched configuration.

This feature has direct implications for the global structure of the phase space. In particular, one might attempt to characterize the dynamics using basin boundary diagrams in the $(R_0,\theta_0)$ plane, distinguishing between collapsing and expanding trajectories. However, \textit{since the line $\theta_0=\pi/2$ universally corresponds to collapsing solutions, the resulting basin structure is highly constrained and does not exhibit a rich fractal boundary.} Consequently, such basin boundary diagrams are not particularly illuminating in the present context, as they do not capture the essential mechanism responsible for the transition to expanding behavior. Instead, the relevant structure is governed by the interplay between angular momentum generation and nonlinear coupling, rather than by a sharp separation of initial conditions.

Another important feature that emerges in our analysis is the role of fractional memory effects. The fractional kernel introduces a time-dependent weighting that modifies both kinetic and potential terms in the effective Lagrangian. In regimes where the evolution occurs far from the memory scale, the kernel becomes effectively constant, explaining why fractional effects are subdominant at early times. Nevertheless, \textit{the presence of the fractional structure qualitatively alters the dynamical system by introducing dissipation-like terms and explicit time dependence}, which become important in determining stability properties and long-term evolution.

The coupled $(R,\theta)$ system also exhibits rich nonlinear behavior, including the emergence of chaos. The Lyapunov analysis reveals that \textit{radial instability is robust across all parameter regimes}, while angular instability is sensitive to both the initial conditions and the fractional parameter. In particular, \textit{fractional memory effects act as an effective stabilizing mechanism for angular motion}, progressively suppressing chaotic behavior as the fractional parameter is decreased. Furthermore, the degree of chaos is strongly controlled by the initial formation time, with early-time configurations displaying strongly chaotic dynamics and later-time configurations exhibiting a transition toward weakly chaotic or near-regular behavior. \textit{Importantly, the onset of chaos is directly linked to the emergence of expanding solutions}, as the amplification of angular motion through nonlinear coupling enables an efficient transfer of energy into the radial sector via the centrifugal term.

These results suggest that the standard picture of cosmic string loop evolution may require revision when additional dynamical ingredients are included. In particular, \textit{the existence of expanding solutions indicates that cosmic string loops may survive longer than previously expected}, potentially enhancing their observational relevance, for instance in the context of gravitational wave production or lensing phenomena. Moreover, \textit{the interplay between angular dynamics and fractional memory introduces a new control mechanism for the stability and fate of loops}, which could have implications for the statistical properties of string networks.

Several open problems naturally arise from the present work. A detailed analysis of gravitational radiation from fractional cosmic string loops would be important to assess how memory effects and angular dynamics modify observable signatures. Extending the present framework to full string network evolution, including loop production and reconnection, would provide a more complete cosmological picture. On a more formal level, it would be interesting to explore whether fractional worldsheet dynamics admits a consistent quantization and how it relates to nonlocal extensions of string theory or quantum gravity. Furthermore, the role of fractional effects in accelerating backgrounds such as de Sitter spacetime deserves further investigation, particularly in light of known results on loop survival and black hole formation. Finally, a more detailed exploration of the phase space structure, beyond simple basin diagnostics, would help clarify the mechanisms underlying the transition between collapsing and expanding trajectories.

In conclusion, the present study demonstrates that fractional extensions of string dynamics, combined with angular degrees of freedom, lead to qualitatively new phenomena in the evolution of cosmic string loops. \textit{The emergence of expanding solutions, the persistence of collapse at $\theta=\pi/2$, and the interplay between chaos and memory effects collectively point toward a richer and more intricate dynamical structure than previously recognized}, opening new directions for both theoretical investigation and phenomenological applications.
\section*{Acknowledgements}

The authors thank the Department of Physics, NIT Silchar, for support and research facilities. We are grateful to our colleagues Juman Sarma and Utkarsh Kumar for useful discussions and helpful comments that improved the manuscript.

\bibliographystyle{JHEP}

\bibliography{ref.bib}

\end{document}

%% file: macros.tex

\newcommand{\plan}[1]{\begin{flushleft}
\gray{\tt *** #1 ***}
\end{flushleft}}

\def\bal#1\eal{\begin{align}#1\end{align}}

\def \bib{\bibitem}
\def\){\right)}
\def\({\left( }
\def\]{\right] }
\def\[{\left[ }

\def\nn{\nonumber}
\def\NO{\nonumber}
\def\nonu{\nonumber \\}
\def\ni{\noindent}

\def\half{\frac{1}{2}}


\newtheorem{definition}{Definition}[section]
\newtheorem{theorem}{Theorem}[section]
\newtheorem{lemma}{Lemma}[section]
\newtheorem{corollary}{Corollary}[section]
\newtheorem{proposition}{Proposition}[section]
\newtheorem{conjecture}{Conjecture}[section]


\def\a{\alpha}
\def\b{\beta}
\def\c{\chi}
\def\d{\delta}
\def\e{\epsilon}
\def\f{\phi}
\def\g{\gamma}
\def\h{\eta}
\def\j{\psi}
\def\k{\kappa}
\def\l{\lambda}
\def\m{\mu}
\def\n{\nu}
\def\om{\omega}
\def\p{\pi}
\def\th{\theta}
\def\r{\rho}
\def\s{\sigma}
\def\t{\tau}
\def\x{\xi}
\def\z{\zeta}
\def\D{\Delta}
\def\F{\Phi}
\def\G{\Gamma}
\def\J{\Psi}
\def\L{\Lambda}
\def\Om{\Omega}
\def\P{\Pi}
\def\Th{\Theta}
\def\S{\Sigma}
\def\U{\Upsilon}
\def\X{\Xi}


\def\ve{\varepsilon}
\def\vr{\varrho}
\def\vs{\varsigma}
\def\vth{\vartheta}
\def\tvf{\tilde{\varphi}}
\def\vf{\varphi}


\def\bba{\bbalpha}
\def\bbk{\bbkappa}
\def\bbg{\bbgamma}
\def\bbd{\bbdelta}
\def\bbs{\bbsigma}


\def\ca{{\cal A}}
\def\cb{{\cal B}}
\def\cc{{\cal C}}
\def\cd{{\cal D}}
\def\ce{{\cal E}}
\def\cf{{\cal F}}
\def\cg{{\cal G}}
\def\ch{{\cal H}}
\def\ci{{\cal I}}
\def\cj{{\cal J}}
\def\ck{{\cal K}}
\def\cl{{\cal L}}
\def\cm{{\cal M}}
\def\cn{{\cal N}}
\def\co{{\cal O}}
\def\cp{{\cal P}}
\def\cq{{\cal Q}}
\def\car{{\cal R}}
\def\cs{{\cal S}}
\def\ct{{\cal T}}
\def\cu{{\cal U}}
\def\cv{{\cal V}}
\def\cw{{\cal W}}
\def\cx{{\cal X}}
\def\cy{{\cal Y}}
\def\cz{{\cal Z}}


\def\bta{\textit{\textbf{a}}}
\def\btb{\textit{\textbf{b}}}
\def\btc{\textit{\textbf{c}}}
\def\btd{\textit{\textbf{d}}}
\def\bte{\textit{\textbf{e}}}
\def\btf{\textit{\textbf{f}}}
\def\btg{\textit{\textbf{g}}}
\def\bth{\textit{\textbf{h}}}
\def\bti{\textit{\textbf{i}}}
\def\btj{\textit{\textbf{j}}}
\def\btk{\textit{\textbf{k}}}
\def\btl{\textit{\textbf{l}}}
\def\btm{\textit{\textbf{m}}}
\def\btn{\textit{\textbf{n}}}
\def\bto{\textit{\textbf{o}}}
\def\btp{\textit{\textbf{p}}}
\def\btq{\textit{\textbf{q}}}
\def\btr{\textit{\textbf{r}}}
\def\bts{\textit{\textbf{s}}}
\def\btt{\textit{\textbf{t}}}
\def\btu{\textit{\textbf{u}}}
\def\btv{\textit{\textbf{v}}}
\def\btw{\textit{\textbf{w}}}
\def\btx{\textit{\textbf{x}}}
\def\bty{\textit{\textbf{y}}}
\def\btz{\textit{\textbf{z}}}

\newcommand{\sgn}{\operatorname{sgn}}

\renewcommand{\bar}{\overline}
\renewcommand{\tilde}{\widetilde}
\renewcommand{\hat}{\widehat}
\renewcommand{\leq}{\leqslant}
\renewcommand{\geq}{\geqslant}
\newcommand{\la}{\left\langle}
\newcommand{\ra}{\right\rangle}
\newcommand{\xp}{x^{+}}
\newcommand{\xm}{x^{-}}

\newcommand{\CC}{\mathbb{C}}
\newcommand{\RR}{\mathbb{R}}
\newcommand{\HH}{\mathbb{H}}
\newcommand{\ZZ}{\mathbb{Z}}
\newcommand{\cA}{\mathcal{A}}
\newcommand{\cB}{\mathcal{B}}
\newcommand{\cC}{\mathcal{C}}
\newcommand{\cD}{\mathcal{D}}
\newcommand{\cE}{\mathcal{E}}
\newcommand{\cF}{\mathcal{F}}
\newcommand{\cG}{\mathcal{G}}
\newcommand{\cH}{\mathcal{H}}
\newcommand{\cI}{\mathcal{I}}
\newcommand{\cJ}{\mathcal{J}}
\newcommand{\cK}{\mathcal{K}}
\newcommand{\cL}{\mathcal{L}}
\newcommand{\cM}{\mathcal{M}}
\newcommand{\cN}{\mathcal{N}}
\newcommand{\cO}{\mathcal{O}}
\newcommand{\cP}{\mathcal{P}}
\newcommand{\cQ}{\mathcal{Q}}
\newcommand{\cR}{\mathcal{R}}
\newcommand{\cS}{\mathcal{S}}
\newcommand{\cT}{\mathcal{T}}
\newcommand{\cU}{\mathcal{U}}
\newcommand{\cV}{\mathcal{V}}
\newcommand{\cW}{\mathcal{W}}
\newcommand{\cX}{\mathcal{X}}
\newcommand{\cY}{\mathcal{Y}}
\newcommand{\cZ}{\mathcal{Z}}


\newcommand{\bra}{\bar{a}}
\newcommand{\brb}{\bar{b}}
\newcommand{\brc}{\bar{c}}
\newcommand{\brd}{\bar{d}}
\newcommand{\bre}{\bar{e}}
\newcommand{\brf}{\bar{f}}
\newcommand{\brg}{\bar{g}}
\newcommand{\brh}{\bar{h}}
\newcommand{\bri}{\bar{i}}
\newcommand{\brj}{\bar{j}}
\newcommand{\brk}{\bar{k}}
\newcommand{\brl}{\bar{l}}
\newcommand{\brm}{\bar{m}}
\newcommand{\brn}{\bar{n}}
\newcommand{\bro}{\bar{o}}
\newcommand{\brp}{\bar{p}}
\newcommand{\brq}{\bar{q}}
\newcommand{\brr}{\bar{r}}
\newcommand{\brs}{\bar{s}}
\newcommand{\brt}{\bar{t}}
\newcommand{\bru}{\bar{u}}
\newcommand{\brv}{\bar{v}}
\newcommand{\brw}{\bar{w}}
\newcommand{\brx}{\bar{x}}
\newcommand{\bry}{\bar{y}}
\newcommand{\brz}{\bar{z}}

\newcommand{\be}{\begin{equation}}
\newcommand{\ee}{\end{equation}}
\newcommand{\bea}{\begin{eqnarray}}
\newcommand{\eea}{\end{eqnarray}}
\newcommand{\bb}{\mathbb}
\newcommand{\ba}{\begin{align}}
\newcommand{\ea}{\end{align}}
\newcommand{\bad}{\begin{aligned}}
\newcommand{\ead}{\end{aligned}}
\newcommand{\nd}{\noindent}
\newcommand{\bsub}{\begin{subequations}}
\newcommand{\esub}{\end{subequations}}
\newcommand{\beqx}{\begin{displaymath}}
\newcommand{\eeqx}{\end{displaymath}}
\newcommand{\bmat}{\left(\begin{array}}
\newcommand{\emat}{\end{array}\right)}
\newcommand*\Laplace{\mathop{}\!\mathbin\bigtriangleup}
\newcommand*\DAlambert{\mathop{}\!\mathbin\Box}



\def\Sc#1{{\hbox{\sc #1}}}      
\def\Sf#1{{\hbox{\sf #1}}}      
\def\mb#1{\mbox{\boldmath $#1$}}
\def\mf#1{\ensuremath{\mathfrak{#1}}} 
\def\bb#1{\ensuremath{\mathbb{#1}}} 


\def\slpa{\slash{\pa}}                         
\def\slin{\SLLash{\in}}                                 
\def\bo{{\raise-.3ex\hbox{\large$\Box$}}}               
\def\cbo{\Sc [}                                         
\def\pa{\partial}                                       
\def\de{\nabla}                                         
\def\dell{\nabla}                                       
\def\su{\sum}                                           
\def\pr{\prod}                                          
\def\iff{\leftrightarrow}                               
\def\conj{{\hbox{\large *}}}                            
\def\ltap{\raisebox{-.4ex}{\rlap{$\sim$}} \raisebox{.4ex}{$<$}}   
\def\gtap{\raisebox{-.4ex}{\rlap{$\sim$}} \raisebox{.4ex}{$>$}}   
\def\face{{\raise.2ex\hbox{$\displaystyle \bigodot$}\mskip-2.2mu \llap {$\ddot
        \smile$}}}                                   
\def\dg{\dagger}                                     
\def\ddg{\ddagger}                                   
\def\trans{\mbox{\scri T}}                           
\def\>{\rangle}                                      
\def\<{\langle}                                      


\def\tx#1{\text{#1}}
\def\sp#1{{}^{#1}}                                   
\def\sb#1{{}_{#1}}                                   
\def\sptx#1{{}^{\rm #1}}                           
\def\sbtx#1{{}_{\rm #1}}                           
\newcommand{\sub}[1]{\phantom{}_{(#1)}\phantom{}}    
\def\oldsl#1{\rlap/#1}                               
\def\slash#1{\rlap{\hbox{$\mskip 1 mu /$}}#1}        
\def\Slash#1{\rlap{\hbox{$\mskip 3 mu /$}}#1}        
\def\SLash#1{\rlap{\hbox{$\mskip 4.5 mu /$}}#1}      
\def\SLLash#1{\rlap{\hbox{$\mskip 6 mu /$}}#1}       
\def\wt#1{\widetilde{#1}}                            
\def\Hat#1{\widehat{#1}}                             
\def\lbar#1{\ensuremath{\overline{#1}}}              
\def\bra#1{\left\langle #1\right|}                   
\def\ket#1{\left| #1\right\rangle}                   
\def\VEV#1{\left\langle #1\right\rangle}             
\def\abs#1{\left| #1\right|}                         
\def\leftrightarrowfill{$\mathsurround=0pt \mathord\leftarrow \mkern-6mu
        \cleaders\hbox{$\mkern-2mu \mathord- \mkern-2mu$}\hfill
        \mkern-6mu \mathord\rightarrow$}        
\def\dvec#1{\vbox{\ialign{##\crcr
        \leftrightarrowfill\crcr\noalign{\kern-1pt\nointerlineskip}
        $\hfil\displaystyle{#1}\hfil$\crcr}}}           
\def\dt#1{{\buildrel {\hbox{\LARGE .}} \over {#1}}}     
\def\dtt#1{{\buildrel \bullet \over {#1}}}              
\def\der#1{{\pa \over \pa {#1}}}                        
\def\fder#1{{\d \over \d {#1}}}                         
\def\tr{{\rm tr \,}}                                    
\def\Tr{{\rm Tr \,}}                                    
\def\diag{{\rm diag \,}}                                
\def\Re{{\rm Re\,}}                                     
\def\Im{{\rm Im\,}}                                     
\def\mrp{\mathrm{p}}

\def\partder#1#2{{\partial #1\over\partial #2}}        
\def\parvar#1#2{{\d #1\over \d #2}}                    
\def\secder#1#2#3{{\partial^2 #1\over\partial #2 \partial #3}}  
\def\on#1#2{\mathop{\null#2}\limits^{#1}}              
\def\bvec#1{\on\leftarrow{#1}}                         
\def\oover#1{\on\circ{#1}}                             


\def\Deq#1{\mbox{$D$=#1}}                               
\def\Neq#1{\mbox{$cn$=#1}}                              
\newcommand{\ampl}[2]{{\cal M}\left( #1 \to #2 \right)} 


\def\NPB#1#2#3{Nucl. Phys. B {\bf #1} (19#2) #3}
\def\PLB#1#2#3{Phys. Lett. B {\bf #1} (19#2) #3}
\def\PLBold#1#2#3{Phys. Lett. {\bf #1}B (19#2) #3}
\def\PRD#1#2#3{Phys. Rev. D {\bf #1} (19#2) #3}
\def\PRL#1#2#3{Phys. Rev. Lett. {\bf #1} (19#2) #3}
\def\PRT#1#2#3{Phys. Rep. {\bf #1} C (19#2) #3}
\def\MODA#1#2#3{Mod. Phys. Lett.  {\bf #1} (19#2) #3}


\def\norder{\raisebox{-.13cm}{\ensuremath{\circ}}\hspace{-.174cm}\raisebox{.13cm}{\ensuremath{\circ}}}
\def\bz{\bar{z}}
\def\bw{\bar{w}}
\def\-{\hphantom{-}}
\newcommand{\dd}{\mbox{d}}
\newcommand{\scr}{\scriptscriptstyle}
\newcommand{\scri}{\scriptsize}
\def\rand#1{\marginpar{\tiny #1}}               
\newcommand{\rstar}{\rand{\bf\large *}}
\newcommand{\rup}{\rand{$\uparrow$}}
\newcommand{\rdown}{\rand{$\downarrow$}}
